\documentclass[pra,showpacs,floatfix]{revtex4}
\usepackage{amsmath}
\usepackage{amssymb}
\usepackage{graphicx}

\newcommand{\beq}{\begin{equation}}
\newcommand{\eeq}{\end{equation}}
\newcommand{\beqa}{\begin{eqnarray}}
\newcommand{\eeqa}{\end{eqnarray}}
\newcommand{\ba}{\begin{array}}
\newcommand{\ea}{\end{array}}
\def\ih{h\kern-0.6em\char"16\kern-0.1em}

\begin{document}

\title{Transition to miscibility in linearly coupled binary dipolar
Bose-Einstein condensates}
\author{Goran Gligori\'c$^1$, Aleksandra Maluckov$^2$, Milutin Stepi\'c$^1$,
Ljup\v co Had\v zievski$^1$, and Boris A. Malomed$^3$}
\affiliation{$^1$ Vin\v ca Institute of Nuclear Sciences, P.O. Box 522,11001 Belgrade,
Serbia \\
$^2$ Faculty of Sciences and Mathematics, University of Ni\v s, P.O. Box
224, 18001 Ni\v s, Serbia \\
$^3$ Department of Physical Electronics, School of Electrical Engineering,
Faculty of Engineering, Tel Aviv University, Tel Aviv 69978, Israel}

\begin{abstract}
We investigate effects of dipole-dipole (DD) interactions on
immiscibility-miscibility transitions (IMTs) in two-component
Bose-Einstein condensates (BECs) trapped in the
harmonic-oscillator (HO) potential, with the components linearly
coupled by a resonant electromagnetic field (accordingly, the
components represent two different spin states of the same atom).
The problem is studied by means of direct numerical simulations.
Different mutual orientations of the dipolar moments in the two
components are considered. It is shown that, in the binary BEC
formed by dipoles with the same orientation and equal magnitudes,
the IMT cannot be induced by the DD interaction alone, being
possible only in the presence of the linear coupling between the
components, while the miscibility threshold is affected by the DD
interactions. However, in the binary condensate with the two
dipolar components polarized in opposite directions, the IMT can
be induced \emph{without} any linear coupling. Further, we
demonstrate that those miscible and immiscible localized states,
formed in the presence of the DD interactions, which are unstable
evolve into robust breathers, which tend to keep the original
miscibility or immiscibility, respectively. An exception is the
case of a very strong DD attraction, when narrow stationary modes
are destroyed by the instability. The binary BEC composed of
co-polarized dipoles with different magnitudes are briefly
considered too.
\end{abstract}

\maketitle

\section{Introduction}

The development of the trapping techniques has made it possible to create
multicomponent Bose-Einstein condensates (BEC) formed by atoms in different
electronic states \cite{trapp}. The multicomponent BEC is an ideal system
for studying phase transitions and the coexistence of different phases,
which are topics of great significance to many areas of physics. The
advantage of the gaseous BEC systems is the possibility to very accurately
describe their dynamics in the framework of the mean-field approximation.
This approach does not apply to multicomponent quantum systems in condensed
states, because their densities are much higher, making the delta-functional
pseudopotential, which is used to describe the local interactions between
atoms in dilute BEC \ by means of the single scattering-length parameter,
inappropriate.

In particular, experiments have been performed in a binary BEC created in $%
^{87}$Rb, which contains atoms in two different hyperfine (spin) states \cite%
{trapp}. The physics of binary BECs gives rise to novel paradigms for
ground-states wave function and excitations on top of them \cite{ground},
\cite{binaryinol}. One generic scenario concerning the binary condensates
leads to an equilibrium state characterized by separation of two immiscible
species in different spatial domains \cite{domain}. In addition, a resonant
electromagnetic spin-flipping field, with a frequency in the range of
several GHz, can induce a linear coupling (interconversion) between the two
species \cite{treca}. This technique opens a way to observe various effects,
such as Josephson oscillations between the two states \cite{peta}, domain
walls \cite{sesta}, ``co-breathing" oscillation modes \cite%
{sedma}, nontopological vortices \cite{osma}, and others.

In the absence of the linear coupling between the components, the superfluid
components formed by atoms in two hyperfine states of $^{87}$Rb, $\left\vert
F=1,m_{F}=-1\right\rangle $ and $\left\vert F=2,m_{F}=+1\right\rangle $, are
immiscible, although being close to the miscibility threshold. On the
contrary, the mixture of $\left\vert F=1,m_{F}=1\right\rangle $ and $%
\left\vert F=2,m_{F}=2\right\rangle $ states is miscible, but the separation
between them can be induced by a gradient of the magnetic field \cite%
{Ketterle}, or by causing them to flow in opposite directions \cite{Engels}.
The dynamics of the separation between the two respective superfluid
components has been studied in detail experimentally \cite{Rb-separation}.

The immiscibility-miscibility transition (IMT) in the binary BEC may
be induced through the Feshbach resonance, which makes it possible
to change the scattering lengths for inter-atomic collisions by
means of an external magnetic field \cite{deseta}, or a properly
tuned coherent optical signal \cite{jedanaesta}. In particular, the
change of the miscibility of $^{85}$Rb and $^{87}$Rb condensates by
means of the Feshbach resonance, which affects the intrinsic
scattering length of the former species, was demonstrated in the
experiments \cite{85-87}.

On the other hand, it has also been shown that the IMT between the
components which represent different spin states of the same atom may be
induced by the linear coupling between them \cite{glavna}. In this way, an
immiscible binary condensate may be converted into a miscible one, using the
spin-flipping electromagnetic wave of a rather small amplitude.

In all the above-mentioned cases, the studies of the binary condensates were
limited to the case of contact inter-particle interactions. On the other
hand, a great deal of attention has been recently drawn to dipolar BECs.
They can be created in gases of ultracold polar molecules \cite{ultra}, in
chromium \cite{chrom}, or in the case when atomic electric dipoles are
induced by the optical pumping of atoms into a Rydberg state. Binary dipolar
BECs are possible too -- in particular, those formed by atoms in two
different Rydberg states \cite{rydberg}. Another realization of a dipolar
mixture is possible in condensates of heteronuclear diatomic molecules
classified as type A according to the Hund's nomenclature \cite{Hund}, in
which the direction of the magnetic moment is correlated (parallel or
antiparallel) with the orientation of the molecular axis \cite{goral}.
Through this mechanism, a binary BEC with \emph{antiparallel} dipolar
components can be created. As shown in Ref. \cite{goral} and in this paper
(see below), binary mixtures of the latter type may feature remarkable
properties.

The goal of this paper is to analyze the IMT in the binary dipolar
BEC formed by the same atoms in different hyperfine states, in the
presence of the linear coupling between the components. It may be
expected that the nonlocal character of the dipole-dipole (DD)
interactions may essentially affect the spatial separation between
the components in the binary condensate. In accordance with this,
the main issues we are addressing here are effects of the DD
interactions on the IMT, including a possibility to \emph{induce}
the IMT by the DD interactions \emph{without} the linear
interconversion between the constituents.

The paper is structured as follows. In section II, equations describing the
two-component dipolar BEC in the harmonic trap, including the linear
coupling between the components, are introduced. This is a system of
one-dimensional (1D) integro-differential equations, with the nonlocal DD
interaction terms set down as in Ref. \cite{goral}. Section III presents
basic results obtained for the binary dipolar mixtures, with different
combinations of orientations and magnitudes of the dipole moments in the two
components. The effect of the DD interactions on the IMT is studied
numerically, for a fixed total number of atoms, $N_{1}+N_{2}=\mathrm{const}$%
. Dynamical properties of localized mixed and separated states are
considered too. The paper is concluded by section IV.

\section{The model}

The quasi-1D dynamical equations for the dipolar BEC trapped in
the harmonic-oscillator (HO) potential are derived starting from
the rescaled Gross-Pitaevskii equation (GPE) with the short-range
and DD interactions,
\begin{equation}
i\frac{\partial }{\partial t}\Psi (\vec{r},t)=\left[ -\frac{1}{2}\nabla ^{2}+%
\frac{1}{2}(\omega _{\rho }^{2}\rho ^{2}+\omega _{z}^{2}z^{2})+\gamma |\Psi (%
\vec{r},t)|^{2}+G\int d\vec{r^{\prime }}\frac{1-3\cos ^{2}\theta }{%
\left\vert \vec{r}-\vec{r^{\prime }}\right\vert ^{3}}|\Psi (\vec{r^{\prime }}%
,t)|^{2}\right] \Psi (\vec{r},t),  \label{3DGPE}
\end{equation}%
following Ref. \cite{lsantos}. In Eq. (\ref{3DGPE}), wave function $\Psi (%
\vec{r},t)$ is normalized to the total number of atoms, $\omega _{\rho }$
and $\omega _{z}$ are normalized trapping frequencies in the transverse and
axial ($z$) direction, $\rho \equiv \sqrt{x^{2}+y^{2}}$, $\theta $ is the
angle between the parallel dipoles and vector $\left( \vec{r}-\vec{r^{\prime
}}\right) $, and coefficients $\gamma $ and $G$ measure, respectively, the
strengths of the contact and DD interactions. The only assumption in the
subsequent derivation is the standard single-mode approximation for the
transversal part of the wave function, i.e., it is substituted by $\Psi
(z,\rho ,t)=\psi (z,t)\phi _{0}(\rho )$, where $\phi _{0}(\rho )=(\sqrt{\pi }%
l)^{-1}\exp (-\rho ^{2}/2l^{2})$, and $l=1/\sqrt{\omega _{\rho }}$ is the
characteristic harmonic-oscillator length which determines the transverse
localization. The averaging in the transverse plane leads to an effective 1D
kernel of the DD interaction \cite{lsantos},
\begin{equation}
V_{\mathrm{DD}}(z)=\frac{2\alpha G}{l^{3}}\left[ 2\sqrt{\tau }-\sqrt{\pi }%
(1+2\tau )\exp {(\tau )}\mathrm{erfc}(\sqrt{\tau })\right] ,  \label{kernel}
\end{equation}%
where $\tau \equiv (z/l)^{2}$, $\alpha =1\,$and $\alpha =-0.5$ for $\theta =0
$ and $\theta =\pi /2$, i.e., respectively, for the polarization of the
dipoles along and perpendicular to axis $z$, and $\mathrm{erfc}$ is the
complementary error function. Being interested in effects of the DD
interactions on the IMT in binary BEC, we will focus on the basic situation,
when the characteristic axial length scale is much larger than the
transverse size, $Z\gg l$. In particular, the experimental setting with
aspect ratio $Z/l\simeq 10$ was realized in the chromium condensates \cite%
{experim1,review}. Under this condition, expression (\ref{kernel}) yields
results which are practically tantamount to those obtained with the use of
the standard form of the DD interaction kernel, $V_{DD}\sim 1/|z|^{3}$,
truncated on the scale of $z\sim l$ (the analysis performed in Ref. \cite%
{Cuevas} for gap solitons in the quasi-1D condensate including the DD
repulsion or attraction and a periodic potential, has demonstrated that the
difference in the results generated by means of the different kernels is
limited to $\lesssim 3\%$). Therefore in the numerical calculations the DD
interaction kernel was taken in the usual regularized (truncated) form.

Proceeding to the consideration of the effectively 1D two-component trapped
BEC, we use a system of nonlinearly coupled GPEs, which include the DD
terms, for the wave functions of the two components, $\psi _{1}(z,t)$ and $%
\psi _{2}(z,t)$. If they represent different hyperfine states of the same
atom, the interconversion, induced by a resonant electromagnetic wave, is
accounted for by linear mixing terms \cite{ist,bitanstabrad}. The scaled
form of the respective GPE system is
\begin{gather}
i\frac{\partial \psi _{1}}{\partial t}=-\frac{1}{2}\frac{\partial ^{2}\psi
_{1}}{\partial z^{2}}+V_{0}z^{2}\psi _{1}+(\gamma _{11}|\psi
_{1}|^{2}+\gamma _{12}|\psi _{2}|^{2})\psi _{1}-\kappa \psi _{2}+  \notag \\
+\psi _{1}\left( G_{11}\int_{-\infty }^{+\infty }\,dz^{\prime }\frac{|\psi
_{1}(z^{\prime })|^{2}}{|z-z^{\prime }|^{3}}+G_{12}\int_{-\infty }^{+\infty
}\,dz^{\prime }\frac{|\psi _{2}(z^{\prime })|^{2}}{|z-z^{\prime }|^{3}}%
\right) ,  \notag \\
i\frac{\partial \psi _{2}}{\partial t}=-\frac{1}{2}\frac{\partial ^{2}\psi
_{2}}{\partial z^{2}}+V_{0}z^{2}\psi _{2}+(\gamma _{21}|\psi
_{1}|^{2}+\gamma _{22}|\psi _{2}|^{2})\psi _{2}-\kappa \psi _{1}+  \notag \\
+\Delta \mu \cdot \psi _{2}+\psi _{2}\left( G_{21}\int_{-\infty }^{+\infty
}\,dz^{\prime }\frac{|\psi _{1}(z^{\prime })|^{2}}{|z-z^{\prime }|^{3}}%
+G_{22}\int_{-\infty }^{+\infty }\,dz^{\prime }\frac{|\psi _{2}(z^{\prime
})|^{2}}{|z-z^{\prime }|^{3}}\right) ,  \label{eq1}
\end{gather}%
where $V_{0}$ is the strength of the axial trap, $\gamma _{11}$, $\gamma
_{22}$ and $\gamma _{12}$, $\gamma _{21}$ are, respectively, the
coefficients accounting for intra-species and inter-species contact
interaction, $\kappa $ is the linear coupling, and $\Delta \mu $ is a
possible chemical-potential difference between the two components, which may
be induced by an external dc magnetic field interacting with the atomic
spins. Further, the DD interaction coefficients, $G_{11},G_{12},G_{21},G_{22}
$, are defined as
\begin{eqnarray}
\left\{ G_{11},G_{12},G_{21}G_{22}\right\}  &=&\left\{ \nu \mu
_{1}^{2},(1-\nu )\mu _{1}\mu _{2},\nu \mu _{1}\mu _{2},\left( 1-\nu \right)
\mu _{2}^{2}\right\}   \notag \\
&&\times \left( \mu _{0}/4\pi \right) (1-3\cos ^{2}{\theta }),  \label{eq2}
\end{eqnarray}%
where $\mu _{0}$ is the magnetic permeability of vacuum, $\mu _{1},\,\mu _{2}
$ are the dipole moments in two BEC components, $\nu \equiv N_{1}/N$ denotes
the ratio of the atom numbers in them, and $N=N_{1}+N_{2}$ is the total
number of atoms. The angle between the system's axis and the dipole
orientations is $\theta $, as before. Below, we focus on the two
above-mentioned basic orientations: parallel ($\theta =0$) and perpendicular
($\theta =\pi /2$), which correspond, respectively to the attractive and
repulsive DD interactions. Two dynamical invariants of Eqs. (\ref{eq1}) are
the total norm,
\begin{equation}
P=\int_{-\infty }^{+\infty }dz\,\left( |\psi _{1}|^{2}+|\psi
_{2}|^{2}\right) ,  \label{norm}
\end{equation}%
and energy (Hamiltonian)
\begin{eqnarray}
E &=&E_{0}+E_{\mathrm{coupl}}+E_{\mathrm{DD}}~,  \notag \\
E_{0} &=&\int_{-\infty }^{+\infty }dz\,\left[ -\frac{1}{2}\left( |\partial
\psi _{1}/\partial z|^{2}+|\partial \psi _{2}/\partial z|^{2}\right)
+V_{0}z^{2}\left( |\psi _{1}|^{2}+|\psi _{2}|^{2}\right) \right]   \notag \\
&+&\int_{-\infty }^{+\infty }dz\,\left[ \frac{1}{2}\left( \gamma _{11}|\psi
_{1}|^{4}+\gamma _{22}|\psi _{2}|^{2}\right) +\gamma _{12}|\psi
_{1}|^{2}|\psi _{2}|^{2}\right] ,  \notag \\
E_{\mathrm{coupl}} &=&-\int_{-\infty }^{+\infty }dz\,\left( \kappa (\psi
_{1}\psi _{2}^{\ast }+\psi _{1}^{\ast }\psi _{2})-\Delta \mu |\psi
_{2}|^{2}\right) ,  \notag \\
E_{\mathrm{DD}} &=&\int_{-\infty }^{+\infty }dz\,\left[ \left( G_{11}|\psi
_{1}|^{2}+G_{21}|\psi _{2}|^{2}\right) \int_{-\infty }^{+\infty }dz^{\prime
}\,\frac{|\psi _{1}(z^{\prime })|^{2}}{|z-z^{\prime }|^{3}}\right] +  \notag
\\
&+&\int_{-\infty }^{+\infty }dz\,\left[ (G_{12}|\psi _{1}|^{2}+G_{22}|\psi
_{2}|^{2})\int_{-\infty }^{+\infty }dz^{\prime }\,\frac{|\psi _{2}(z^{\prime
})|^{2}}{|z-z^{\prime }|^{3}}\right] .  \label{hamiltonian}
\end{eqnarray}

Stationary solutions are looked for as $\psi _{1,2}(z,t)=\phi _{1,2}(z)\exp {%
(-i\mu \,t)}$, where $\mu $ is the chemical potential (in the presence of
the linear coupling, the chemical potential of both components must be the
same \cite{glavna}), and real functions obey the following
integro-differential equations:
\begin{eqnarray}
\mu \phi _{1} &=&-\frac{1}{2}\frac{d^{2}\phi _{1}}{dz^{2}}+V_{0}z^{2}\phi
_{1}+(\gamma _{11}|\phi _{1}|^{2}+\gamma _{12}|\phi _{2}|^{2})\phi
_{1}-\kappa \phi _{2}+  \notag \\
&+&\phi _{1}\left( G_{11}\int_{-\infty }^{+\infty }\,dz^{\prime }\frac{|\phi
_{1}(z^{\prime })|^{2}}{|z-z^{\prime }|^{3}}+G_{12}\int_{-\infty }^{+\infty
}\,dz^{\prime }\frac{|\phi _{2}(z^{\prime })|^{2}}{|z-z^{\prime }|^{3}}%
\right)   \notag \\
\mu \phi _{2} &=&-\frac{1}{2}\frac{d^{2}\phi _{2}}{dz^{2}}+V_{0}z^{2}\phi
_{2}+(\gamma _{21}|\phi _{1}|^{2}+\gamma _{22}|\phi _{2}|^{2})\phi
_{2}-\kappa \phi _{1}+  \notag \\
&+&\phi _{2}\left( G_{21}\int_{-\infty }^{+\infty }\,dz^{\prime }\frac{|\phi
_{1}(z^{\prime })|^{2}}{|z-z^{\prime }|^{3}}+G_{22}\int_{-\infty }^{+\infty
}\,dz^{\prime }\frac{|\phi _{2}(z^{\prime })|^{2}}{|z-z^{\prime }|^{3}}%
\right) +\Delta \mu \cdot \phi _{2}.  \label{stationary}
\end{eqnarray}

Properties of the binary BEC described by this model depend on a large
number of parameters, including the number of particles in each component
and the magnitudes and orientations of the dipoles. In the following, the
results are presented for $\gamma _{11}=\gamma _{22}=10,\,\gamma
_{12}=\gamma _{21}=20,\,\Delta \mu =0,\,V_{0}=0.5$, and equal numbers of
atoms in both components, i.e., $\nu =0.5$. As for the DD terms, we assume
that the dipole moments of the two components are oriented in the same
direction, with all the DD strengths being equal, $%
G_{11}=G_{22}=G_{12}=G_{21}\equiv \Gamma $ (case I), or with
different strengths, $G_{11}=\Gamma ,\,G_{12}=G_{21}=1.1\Gamma$,
and $G_{22}=1.21\Gamma $ (case II). As seen from Eqs. (\ref{eq2}),
cases I and II correspond,
respectively, to equal magnitudes of the dipole moments, $\mu _{1}=\mu _{2}$%
, and unequal ones, $\mu _{2}=1.1\mu _{1}$. The case of the BEC composed of
two identical components polarized in opposite directions, $%
G_{11}=G_{22}=-G_{12}=-G_{21}=\Gamma $ (case III) is considered too. The
latter one corresponds to the above-mentioned setting composed of type-A
molecules, in terms of the Hund nomenclature, under the action of the
magnetic field, which determines the orientation of the magnetic moments,
while the \textit{electric} dipolar moments can be parallel or antiparallel
to the direction of the applied magnetic field \cite{goral}.

In the absence of the DD interactions and linear coupling, Eqs. (\ref{eq1})
or (\ref{stationary}) can be solved, under specific conditions, analytically
by means the inverse scattering technique \cite{ist}, and numerically in the
general case \cite{numoutdd}. In particular, both the analytical and
numerical solutions produce solitons. On the other hand, the localized
miscible and immiscible states were found in the condensate with two
components coupled by the linear interconversion. The IMT in the binary
condensate can be controlled by the strength of the linear driving \cite%
{glavna}. The critical value of the linear-coupling parameter for the IMT
was estimated, by means of the variational approximation, to be $\kappa _{%
\mathrm{cr}}=(\gamma _{12}-\gamma _{11})/2$, which has been
confirmed by the numerical simulations. In this work, we use
numerical methods to  find fundamental solutions of Eq.
(\ref{eq1}) for the binary dipolar BEC, and explore effects of the
DD interactions on the IMT in this system.

\section{Immiscibility-miscibility transitions}


In this section, we present results corresponding to the fixed norm
(rescaled number of atoms), $P\approx 4$, a free parameter being the linear
coupling, $\kappa $. As mentioned above, the parameters of contact
interactions and trapping potential are the same as in Ref. \cite{glavna}: $%
\gamma _{11}=\gamma _{22}=\gamma _{12}/2=\gamma _{21}/2=10$ and $V_{0}=0.5$,
which correspond to the strong immiscibility, in the absence of the linear
coupling and DD interactions. This may be realized so that the two species
do not mix because their mutual repulsion is stronger than the intrinsic
repulsion between atoms belonging to the same species.

As concerns the parameters of the DD interactions, we will consider the
following characteristic cases, I, II, and III (as mentioned above): $%
G_{11}=G_{12}=G_{21}=G_{22}\equiv \Gamma $, or $G_{11}\equiv \Gamma
,\,G_{12}=G_{21}=1.1\Gamma ,\,G_{22}=1.21\Gamma $, or $%
G_{11}=-G_{12}=-G_{21}=G_{22}\equiv \Gamma $. Recall that $\Gamma <0$ and $%
\Gamma >0$ correspond to the attractive and repulsive DD interactions,
respectively. Stationary states were obtained by simulations of Eqs. (\ref%
{eq1}) in imaginary time. In the absence of the DD interactions, slightly
perturbed stationary states maintain themselves (in real time) as breathing
localized modes with the conserved norm \cite{glavna}.

\subsubsection{Equal dipole moments in both components}

\begin{figure}[tbp]
\center\includegraphics [width=5.7cm]{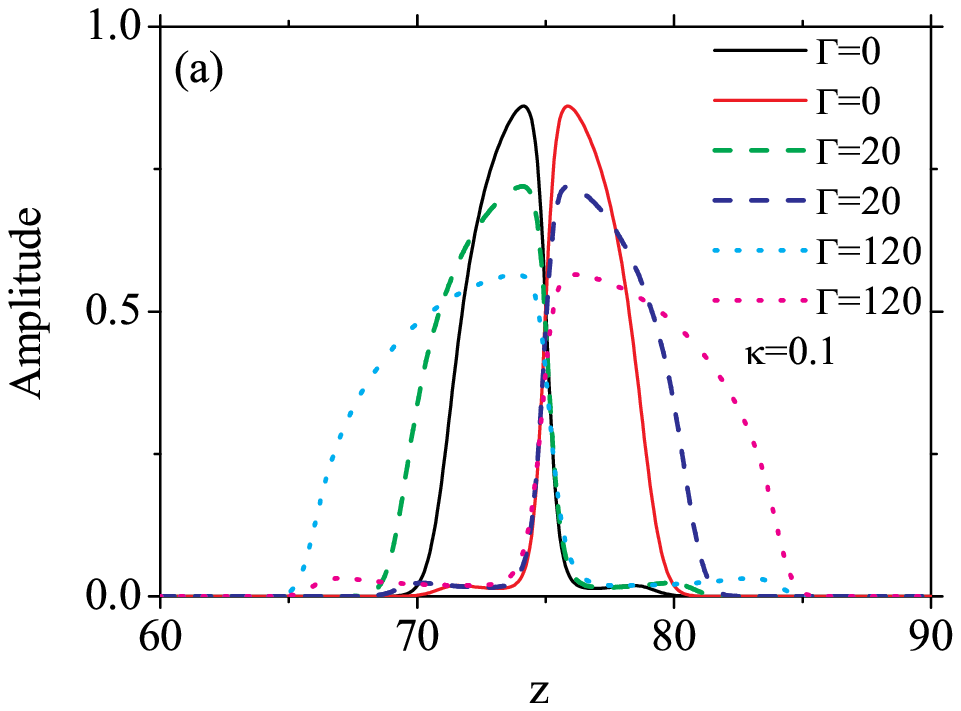}%
\includegraphics
[width=5.7cm]{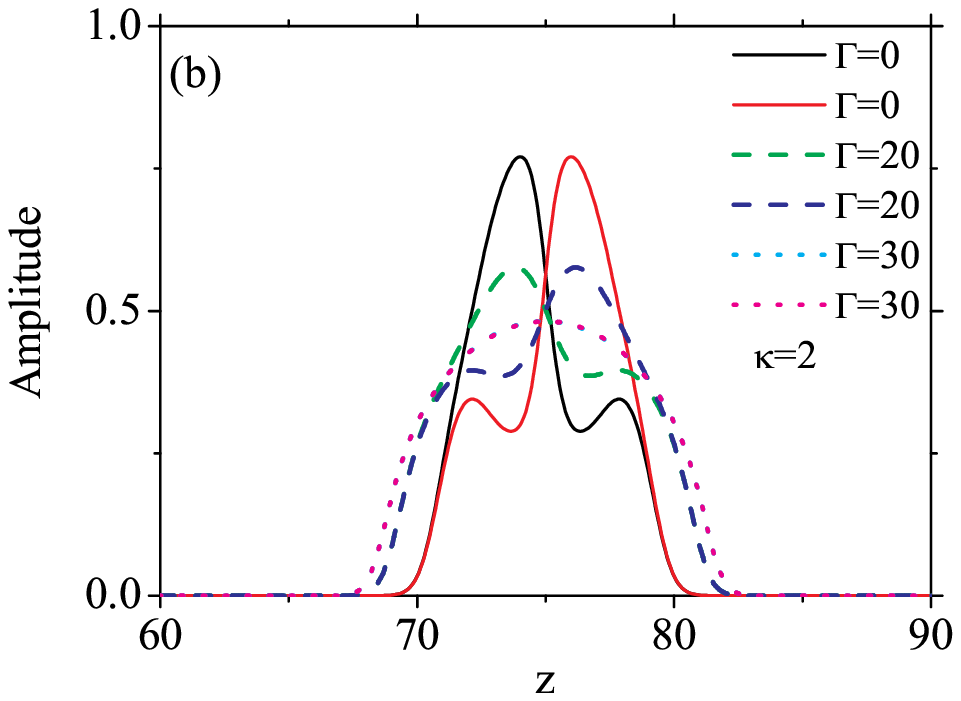} \includegraphics [width=5.7cm]{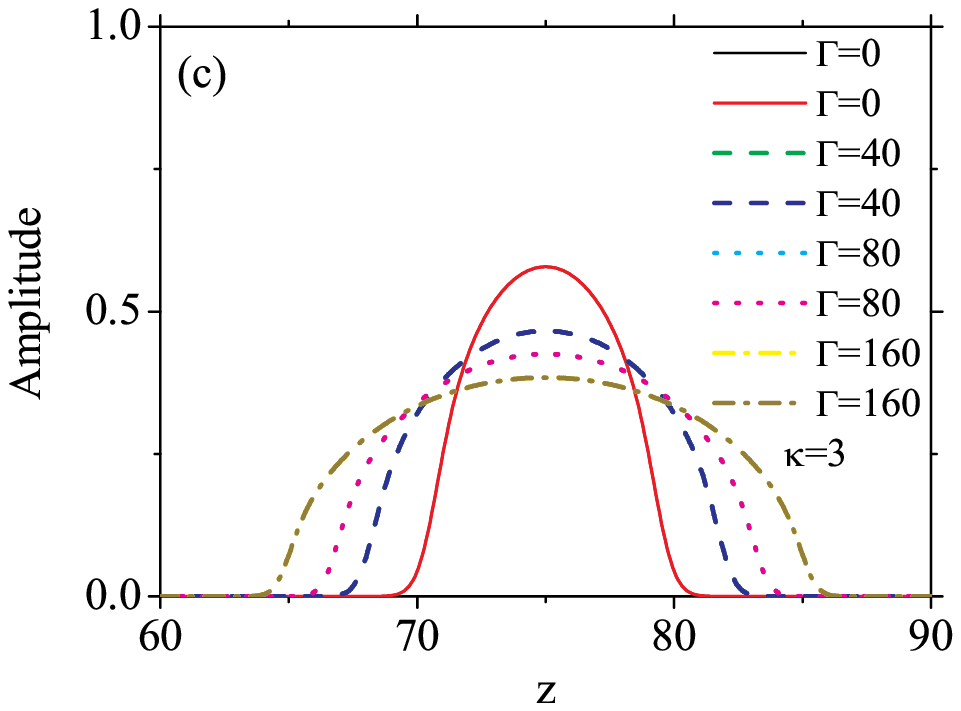}
\caption{(Color online) The amplitude profiles of stationary
states in the
binary BEC for different values of the linear-coupling constant: (a) $%
\protect\kappa =0.1$, (b) $\protect\kappa =2$, and (c)
$\protect\kappa =3$. In each case, the profiles are displayed for
the varying strength of the repulsive DD interaction,
$G_{11}=G_{22}=G_{21}=G_{12}\equiv \Gamma $. In this and other
figures, parameters $\protect\gamma _{mn}\ $($m,n=1,2$),
which characterize the contact repulsion, are $2\protect\gamma _{11}=2%
\protect\gamma _{22}=\protect\gamma _{12}=\protect\gamma _{21}=20$. }
\label{fig1}
\end{figure}

In the absence of the linear coupling between the components, or
with a very weak linear coupling, the addition of the arbitrary
strong repulsive DD interaction to an initially immiscible state
of the binary system cannot induce a transition to miscibility,
see Fig. \ref{fig1}(a). However, the increase of the repulsive
DD-interaction strength in the presence of a somewhat stronger
linear coupling, which still corresponds to the immiscible
phase in the case with only contact interaction, $\kappa <\kappa _{\mathrm{cr%
}}$ (see Ref. \cite{glavna}), can stimulate the IMT, as shown in \ref{fig1}%
(b). In the presence of the repulsive DD interactions, the miscible state
generated by the linear interconversion remains miscible, but the nonlocal
interaction makes the amplitude of the trapped mixed state smaller, and its
width larger, as seen in Fig. \ref{fig1}(c). These results may be summarized
by stating that the repulsive DD interaction tends to mix the separated
components, but it can do this only in the combination with the linear
interconversion between them, while the interplay of the DD repulsion with
the contact nonlinear interactions is less important, in terms of the
qualitative description.

The stability of the localized states was checked by direct simulations of
their perturbed evolution in the real time. It was found that both the
miscible and immiscible states are stable, although the former ones are more
sensitive to asymmetric perturbations than their immiscible counterparts
with the same norm. Generally, the perturbations give rise to small
persistent intrinsic vibrations in the trapped states (not shown here).

In the presence of the \emph{attractive} DD interaction, the components that
were separated at $\kappa =0.1$, in the absence of the DD interactions,
remain separated. Naturally, with the increase of the DD attraction
strength, the localized states become narrower and taller, see Fig. \ref%
{fig2}. However, localized states cannot be found for an arbitrarily large
strength of the DD attraction (an explanation to this fact is given below).
In other words, there is an upper limit of the strength, beyond which the
attractive DD forces cannot support localized modes. The latter conclusion
is consistent with the real-time simulations, which confirm that the BEC
components stay separated, unless the DD attraction is so strong that it
destroys both components. An example of the evolution of the initially
separated state in the presence of the strong attractive DD interaction, but
yet below the destruction threshold, is shown in Fig. \ref{fig3} for $\Gamma
=-10$. Although the central part of the mode stays localized, the mode
radiates a significant part of its norm.

\begin{figure}[tbp]
\center\includegraphics [width=5.7cm]{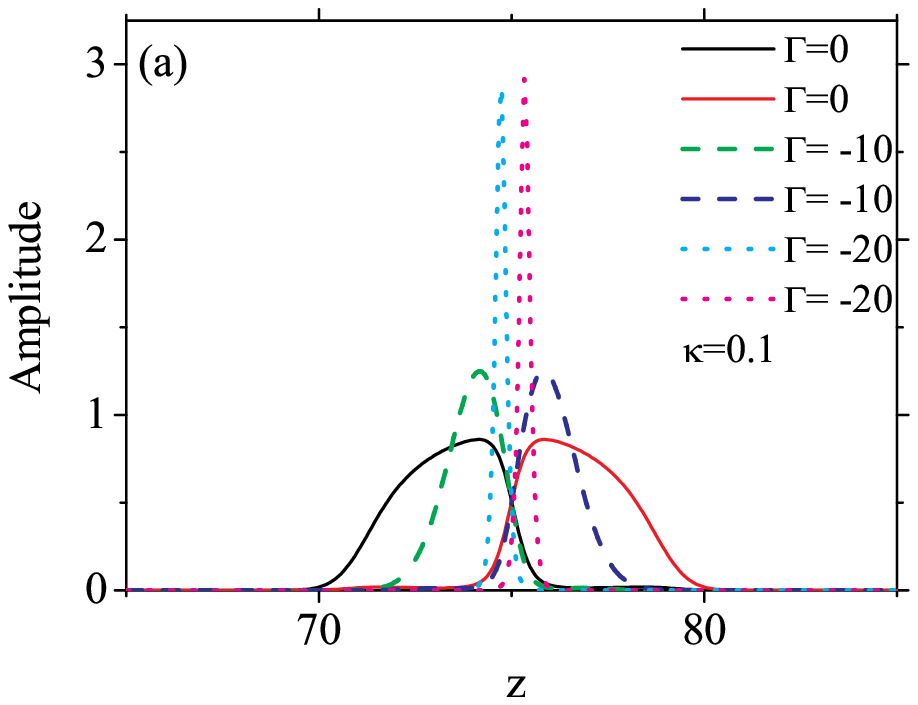}%
\includegraphics
[width=5.7cm]{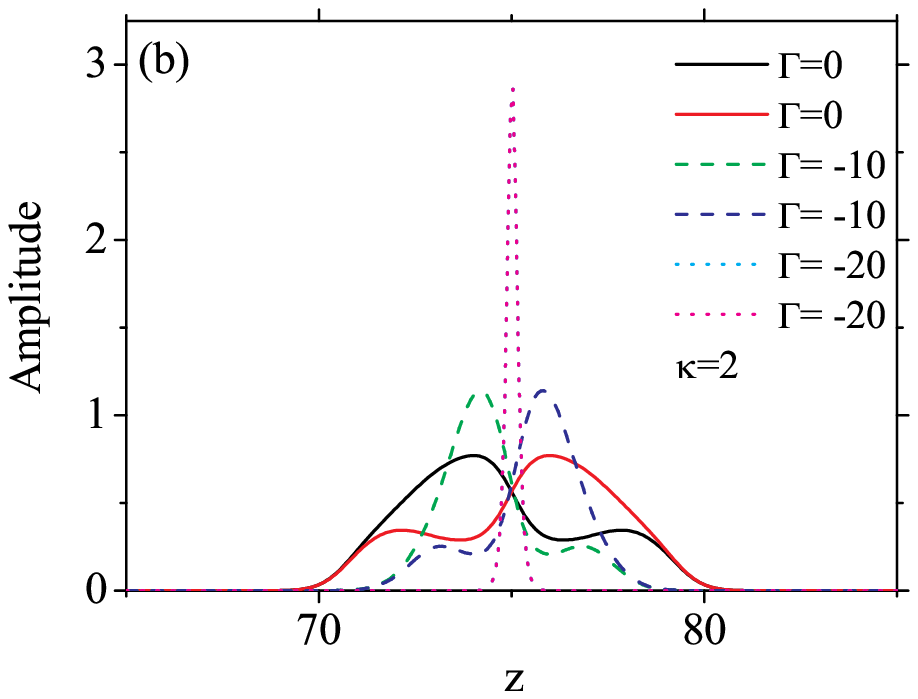} \includegraphics [width=5.7cm]{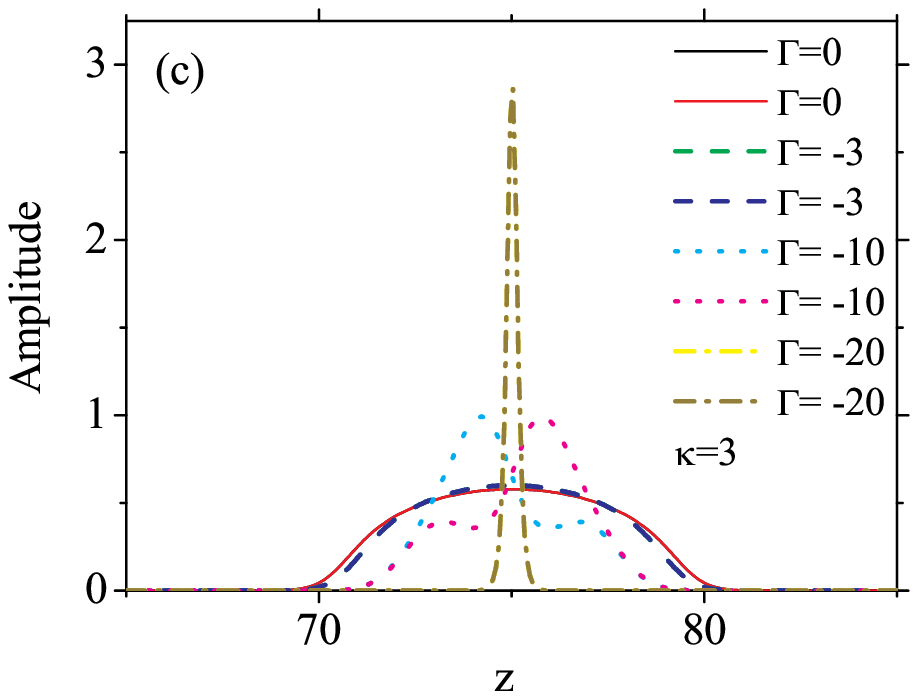}
\caption{(Color online) The amplitude profiles of stationary
states of the
binary BEC for different values of $\protect\kappa $: (a) $\protect\kappa %
=0.1$, (b) $\protect\kappa =2$, and (c) $\protect\kappa =3$, in
the presence of the attractive DD interaction with $\Gamma \equiv
G_{11}=G_{22}=G_{21}=G_{12}$. The values of $\Gamma <0$ are
indicated in the plots.} \label{fig2}
\end{figure}

\begin{figure}[tbp]
\center\includegraphics [width=5.7cm]{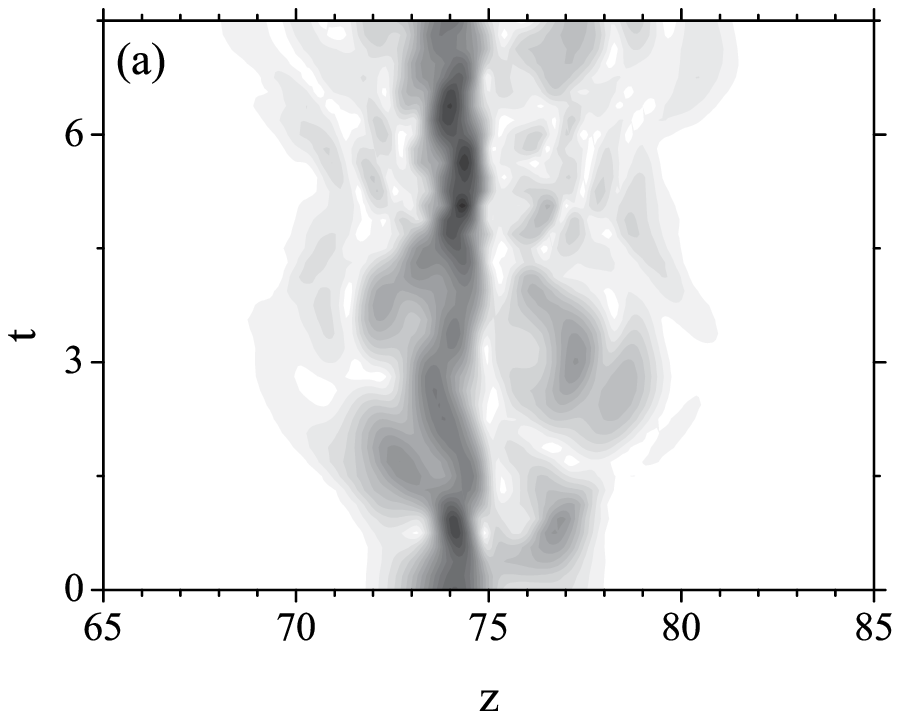}\includegraphics
[width=5.7cm]{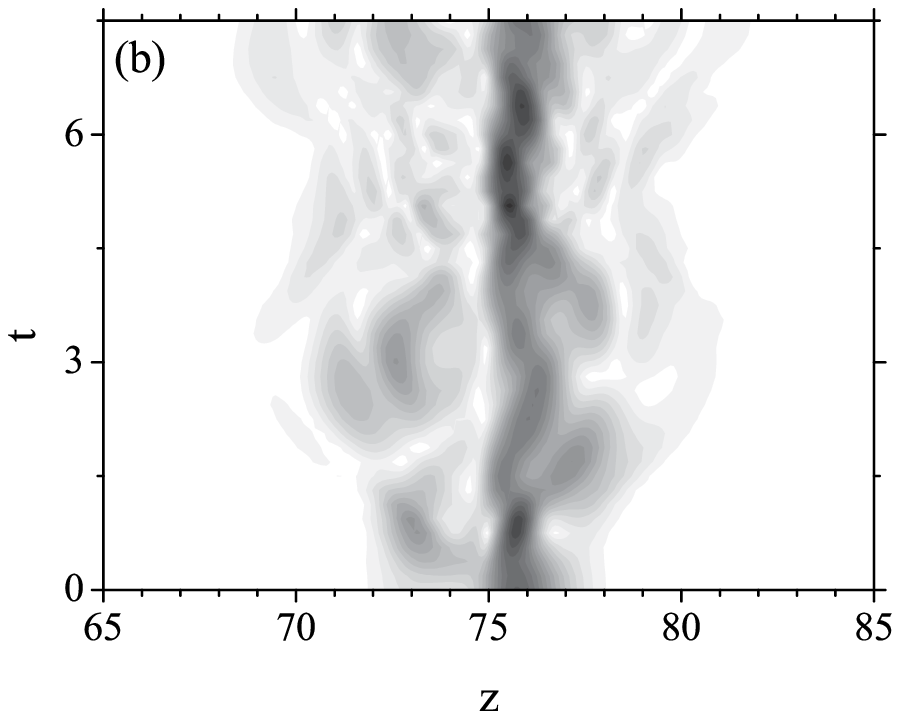} \includegraphics [width=5.7cm]{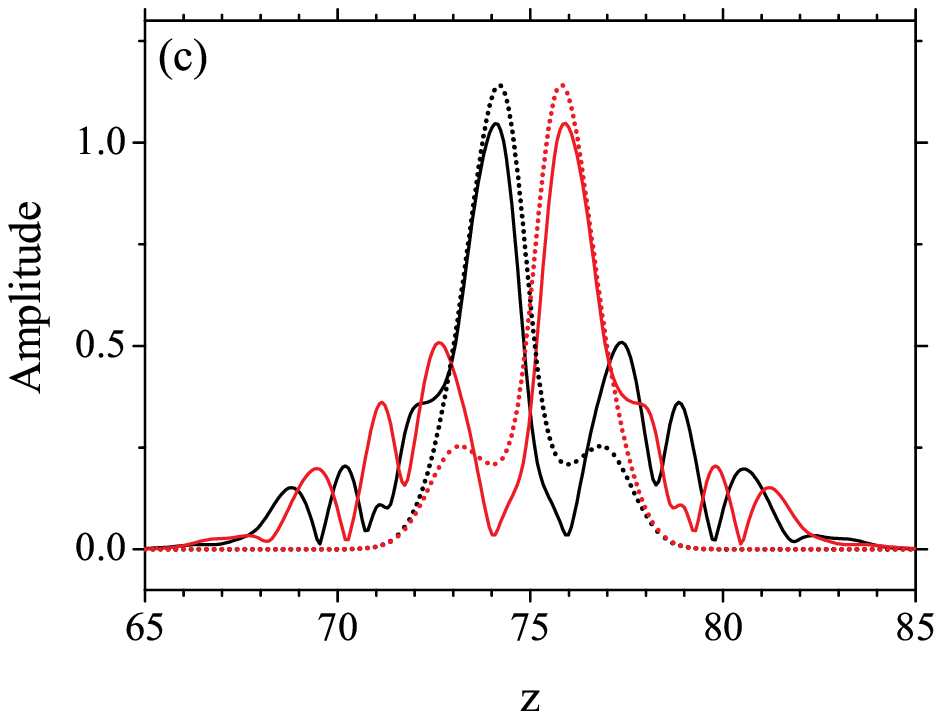}
\caption{(Color online) The amplitude profiles versus time for a
slightly perturbed immiscible state with $\protect\kappa
=2,\,\Gamma =-10$. Panels (a) and (b) display the evolution of the
two components. The initial (dotted and dotted-dashed) profile and
the one at $t=7.5$ (solid and dashed curves) are shown in (c).}
\label{fig3}
\end{figure}

\begin{figure}[tbp]
\center\includegraphics [width=6cm]{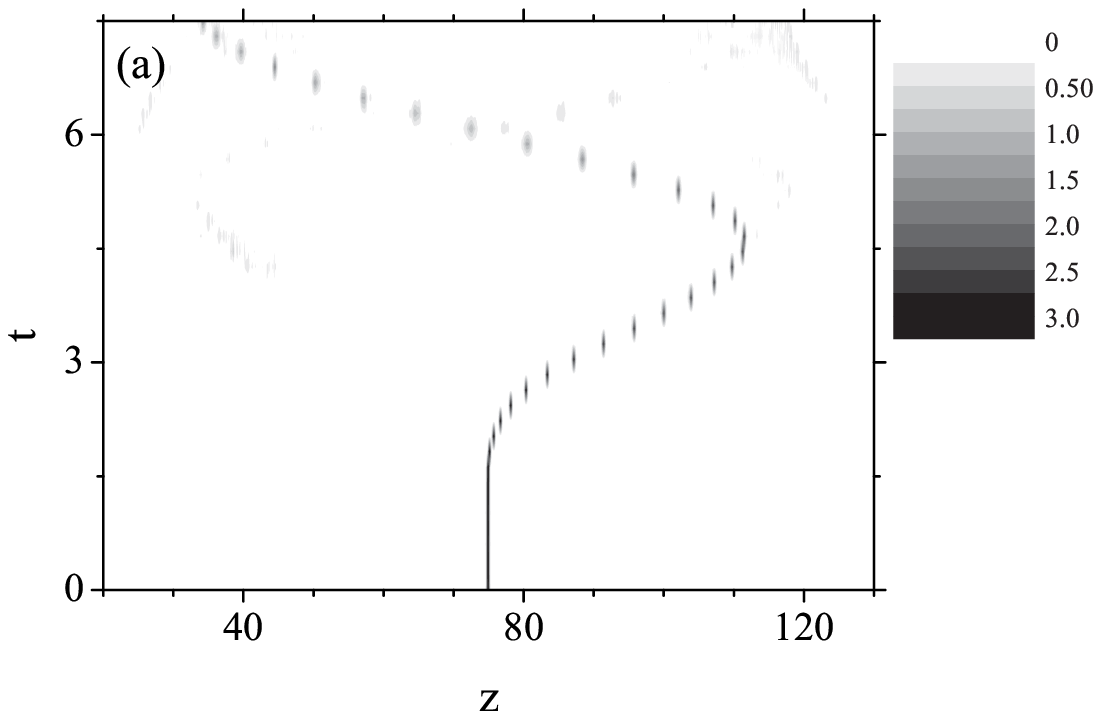}\includegraphics
[width=6cm]{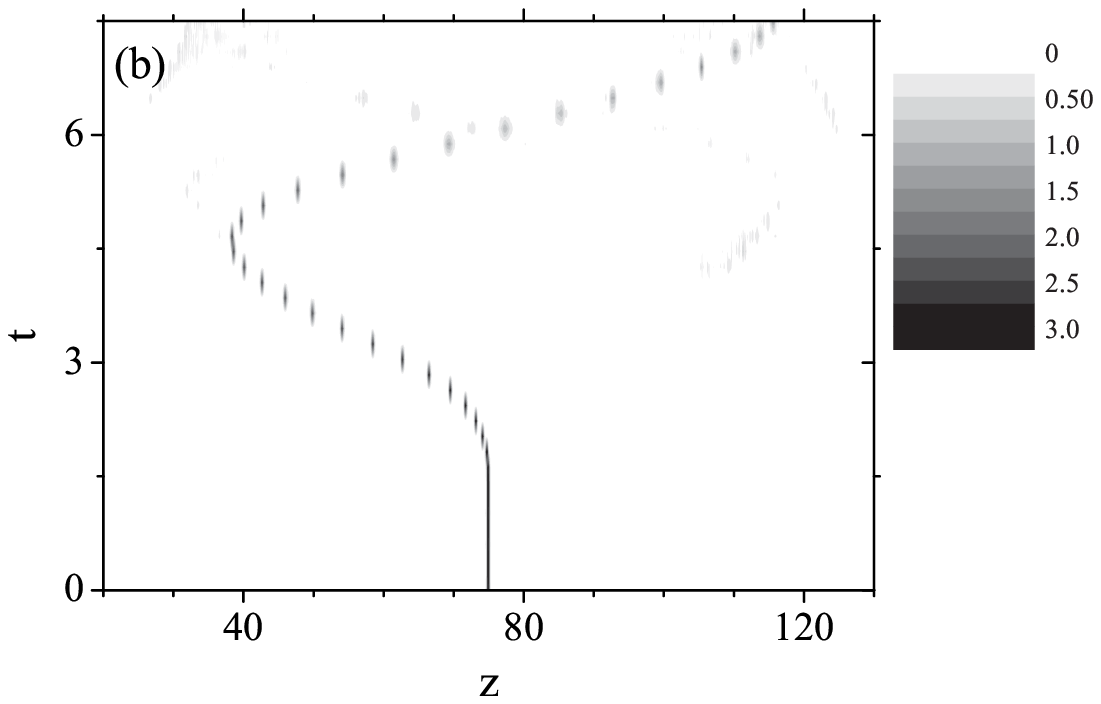} \includegraphics [width=5cm]{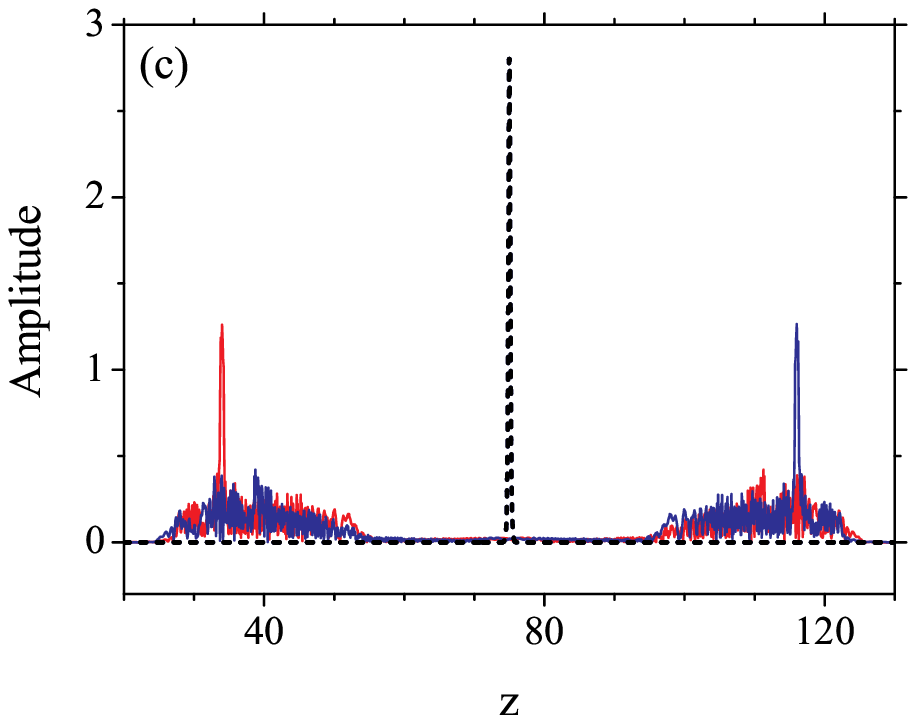}
\caption{(Color online) The amplitude profiles versus time for a
slightly perturbed mixed state with $\protect\kappa =2,\,\Gamma
=-20.$Panels (a) and (b) show the evolution of the two components.
The initial very narrow profile (dotted black curves) and the one
at $t=7.5$ (continuous blue and red curves) are plotted in (c).}
\label{fig4}
\end{figure}

In the parameter region near the IMT induced by the linear coupling in the
absence of the DD interactions [for example, at $\kappa =2$, see Fig. \ref%
{fig2}(b)], the increase of the strength of the attractive DD interaction
again makes the immiscible localized mode narrower and taller. Very strong
DD attraction may result in a miscible state with a very high amplitude and
very narrow shape. However, such states are \emph{unstable} and disintegrate
with time. An example of the instability is displayed in Fig. \ref{fig4}.

In the region of $\kappa $ slightly above $\kappa _{\mathrm{cr}}$, i.e., in
the states which are miscible in the absence of \ the DD interactions, the
presence of the weak DD\ attraction may lead to a transition back to the
immiscible state, which is unstable. In the rest of the region with $\kappa
>\kappa _{\mathrm{cr}}$, the increase of the attractive DD interaction leads
to taller and narrower miscible localized states, see Fig. \ref{fig2} (c).
These newly formed miscible states are also unstable to small perturbations.
Localized states cannot be formed if the DD attraction is too strong,
similar to what was said above for smaller $\kappa $.

The nonexistence of localized states in binary BEC for arbitrary values of $%
\kappa $ in the presence of very strong attractive DD interactions is a
consequence of the singularity in the 1D condensate model, which manifests
itself when the characteristic length of the DD interaction becomes
comparable to the inter-particle distance \cite{goral}. This is related to
the singularity in the single-component BEC, in the case when the negative
pressure caused by the inter-particle attraction is not compensated by the
quantum pressure in the trapping potential \cite{rydberg}, \cite{goral}.

\begin{figure}[tbp]
\center\includegraphics [width=13cm]{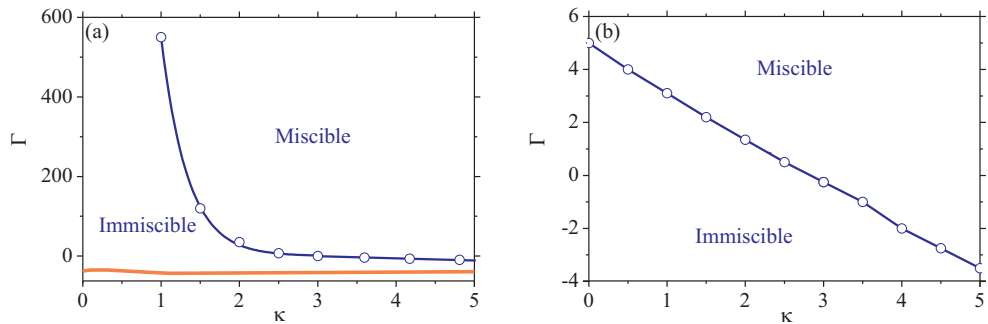}
\caption{(Color online) The blue lines with circles show the
immiscibility-miscibility transition curves in the parameter space, ($\Gamma
$,\thinspace\ $\protect\kappa $), produced by the numerical calculations for
cases I (co-polarized components, panel a) and III (counter-polarized
components, panel b). The areas of miscible and immiscible states (above and
below the transition lines with symbols) are labeled accordingly. Below the
red thick line, localized states in the binary BEC with co-polarized dipolar
components (case I) were not found. }
\label{fig5}
\end{figure}

Thus, the DD interactions can effect the IMT in the dipolar BEC mixture,
with the components polarized in the same direction (and with equal
magnitudes of the magnetic moments), which are coupled by the linear
interconversion. This is a direct consequence of the nonlocality of the DD
interactions. However, the DD interactions of either sign (repulsive or
attractive), by themselves, cannot induce the IMT, as seen in Fig. \ref{fig5}%
(a). On the other hand, the repulsive and attractive DD
interactions' shift of the equilibrium in favor of the miscibility
and immiscibility, respectively.

\begin{figure}[tbp]
\center\includegraphics [width=8cm]{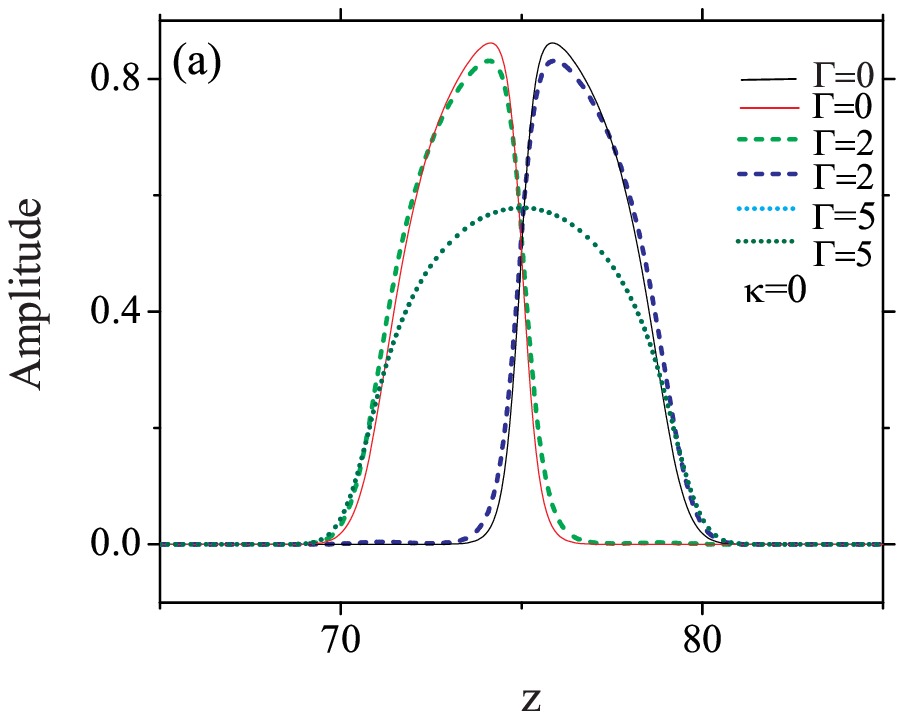}%
\includegraphics
[width=8cm]{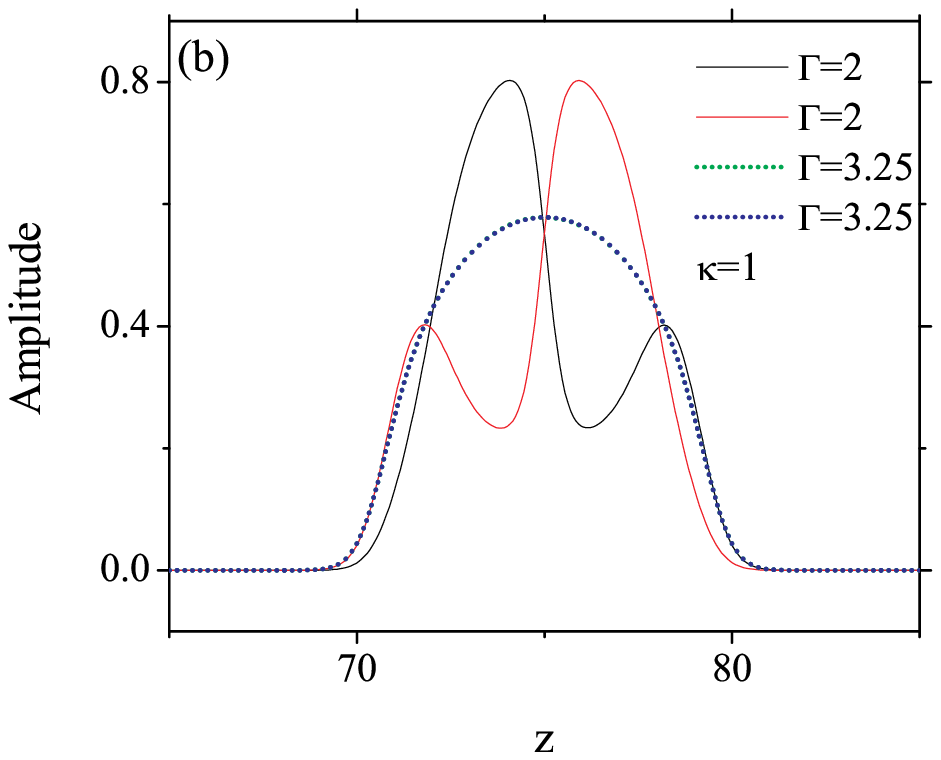} \caption{(Color online) Amplitude profiles
of stationary states for the two-component dipolar mixture with
counter-polarized components: (a) in the
absence of the linear coupling ($\protect\kappa =0$) and (b) for $\protect%
\kappa =1$. Parameters of the DD interaction are $G_{11}=G_{22}=\Gamma
,\,G_{21}=G_{12}=-\Gamma $, with values of $\Gamma $ indicated in the plot.
The transition to the miscibility with the increase of $\Gamma $ is evident.}
\label{fig6}
\end{figure}

A transition to the miscibility may be induced by the DD
interactions (in the absence the linear interconversion) if the
two dipolar components are polarized in the \emph{opposite}
directions, which was defined above as case III, with $\mu
_{1}=-\mu _{2}$ in Eqs. (\ref{eq2}). Recall that such a situation
can be realized for molecules of the Hund's type A in the external
magnetic field \cite{goral}. The possibility to induce the IMT by
the DD interaction in this case is illustrated by Fig. \ref{fig6}
(a), which corresponds to the model without the linear coupling.

\begin{figure}[tbp]
\center\includegraphics [width=5.7cm]{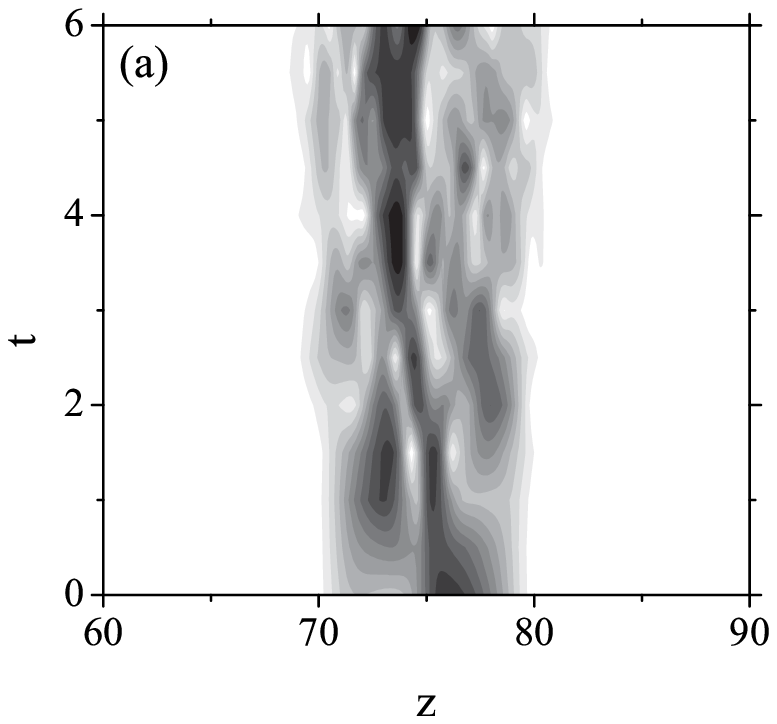}%
\includegraphics
[width=5.7cm]{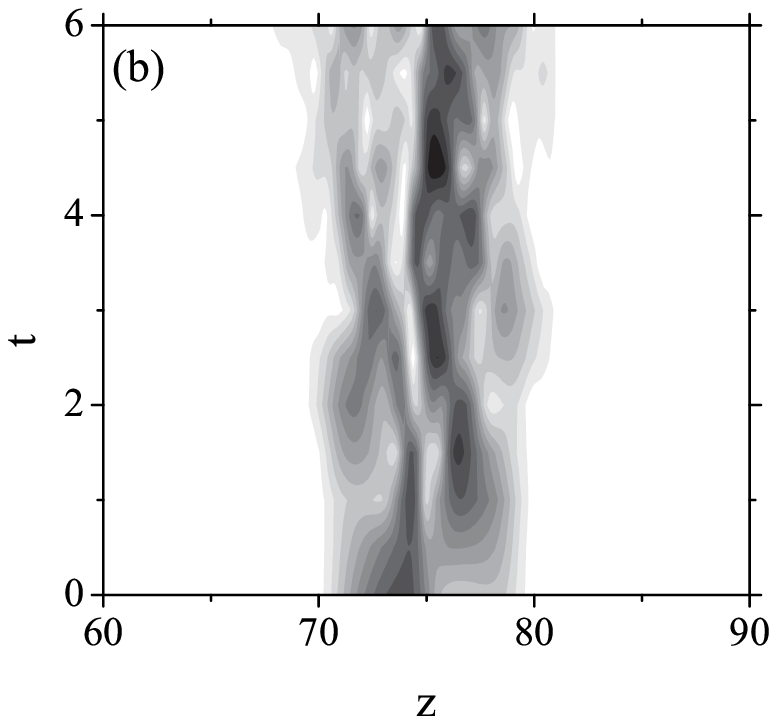}\includegraphics [width=5.7cm]{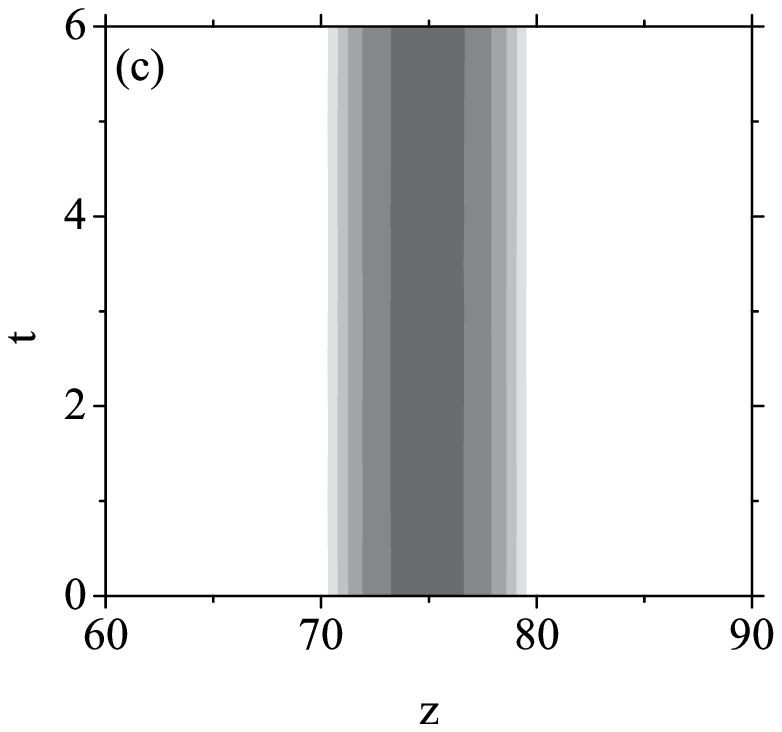}
\caption{2D plots of the evolution in case III (counter-polarized
components) with $\protect\kappa =1$: (a) and (b) the evolution of
the two components initialized in the immiscible state for $\Gamma
=2$; (c) the evolution of a stable miscible state for $\Gamma
=3.25$, cf. Fig. \protect\ref{fig6}(b). The immiscible states
evolve as localized structures with complex intrinsic patterns,
while the miscible state evolves as a regular breather.}
\label{fig7}
\end{figure}

The numerically found IMT curve for case III (the counter-polarized
components) in the parameter space of ($\Gamma $, $\kappa $) is displayed in
Fig. \ref{fig5}(b). In this case, the addition of the linear interconversion
shifts the transition to the miscibility to a lower strength of the DD
interaction. Figure \ref{fig7} displays examples of the evolution of
immiscible and miscible states for $\kappa =1$ which are presented in Fig. %
\ref{fig6}(b). The immiscible state evolves as a localized mode with a
complex intrinsic structure, while the miscible state is only slightly
perturbed, evolving into a regular breather. Note that, in case III, with
the opposite polarizations of the dipoles in the two components, i.e., $%
G_{11}=G_{22}=\Gamma <0,\,G_{12}=G_{21}=\Gamma >0$, the IMT to a stable
miscible state cannot be induced by the DD interactions in the absence of
the linear interconversion, see Fig. \ref{fig5} (b). In this case, only a
narrow strongly pinned miscible state is formed, which is unstable.

In fact, the configuration with the counter-polarized components considered
here implies that, in both of them, the moments are directed along the
spatial axis. The other case, with the counter-polarized components oriented
perpendicular to the axis, is possible too, in which case the DD interaction
is attractive between the components and repulsive inside each component.

To summarize the analysis of cases I and III, we have found that
the repulsive DD interactions affect the IMT, but only in the
presence of the linear coupling in the case I-- namely, the curve
with circles in Fig. \ref{fig5}(a) shows the critical values of
$\kappa $ which are lower than their counterparts found in the
absence of the DD interactions. In this case, both the miscible
and immiscible stationary modes remain stable under the action of
the DD repulsion. The weak attractive DD interactions may convert
a
miscible state near the IMT threshold ($\kappa \approx \kappa _{\mathrm{cr}}$%
) into an \emph{unstable} immiscible one. At larger values of $\kappa
>\kappa _{\mathrm{cr}}$, the increase of the attractive DD interaction
causes the formation of tightly pinned miscible states, which are
strongly unstable. There is a limiting value of the strength of
the DD attraction, above which localized states cannot exist in
the binary BEC. Finally, for the condensate with counter-polarized
components (case III),
the numerical results reveal the existence of the single IMT point, $\Gamma $%
, which moves to smaller values with the increase of $\kappa $, see Fig. \ref%
{fig5} (b).

\subsubsection{Unequal magnitudes of the dipole moments in the two components%
}

In the model describing binary BEC formed of atoms with equal dipole
moments, which was considered above, it is obvious that all the states are
spatially symmetric, in terms of the total density of both species (before
and after the IMT, and for both the co- and counter-polarized components).
On the other hand, in the binary system with different dipole moments of the
two species an issue is not only the IMT itself, but also the equilibrium
partition of atoms between the species [cf. Ref. \cite{glavna}, where an
asymmetry between the species was introduced through the difference in the
chemical potentials, $\Delta \mu $ in Eqs. (\ref{eq1}).

\begin{figure}[tbp]
\center\includegraphics [width=8cm]{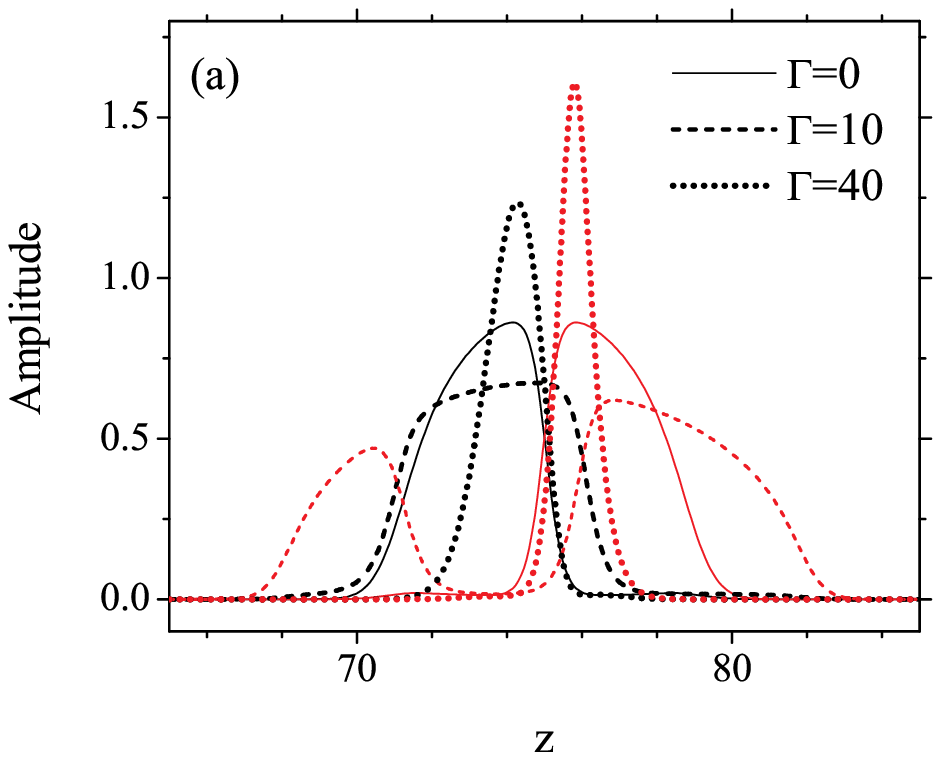}\includegraphics
[width=8cm]{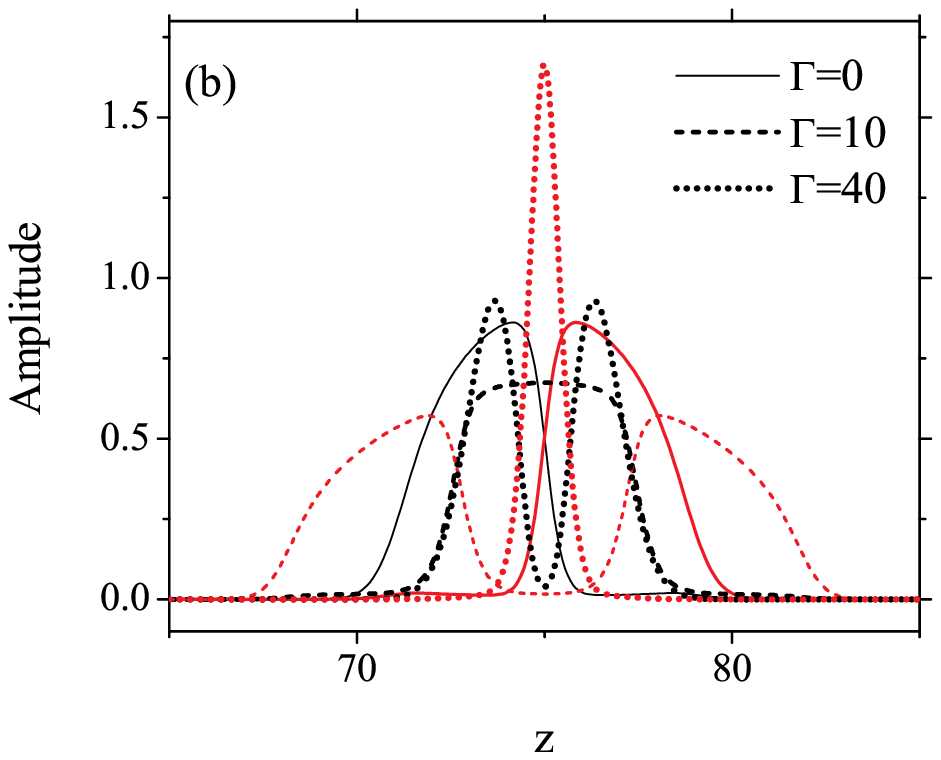} \includegraphics
[width=8cm]{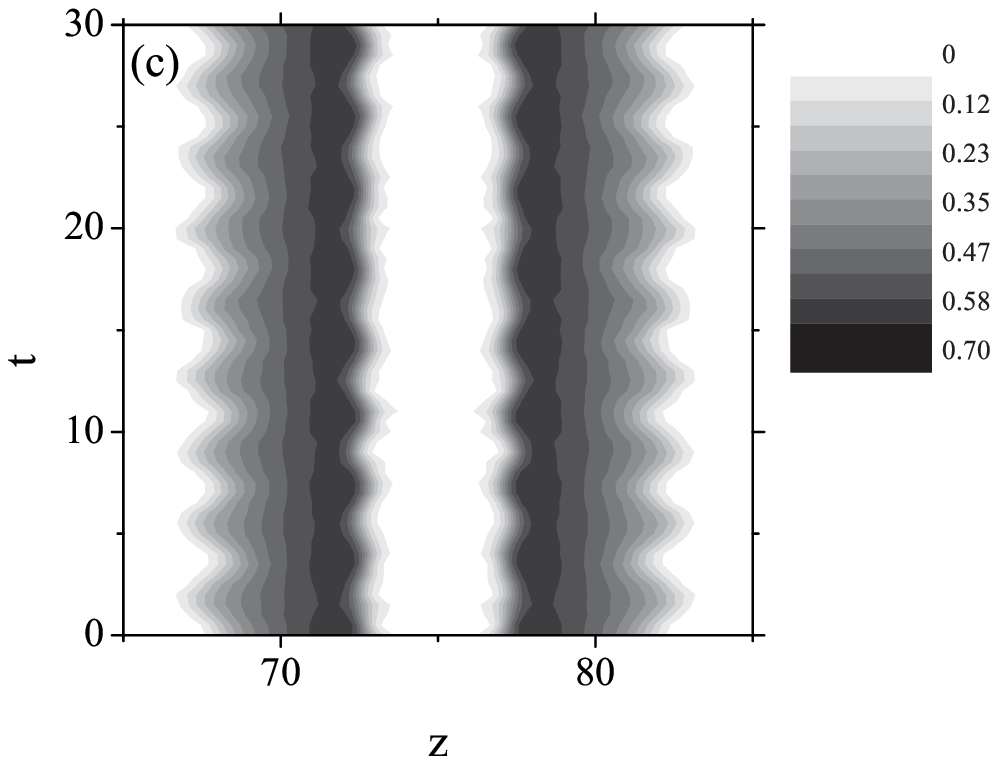}\includegraphics [width=8cm]{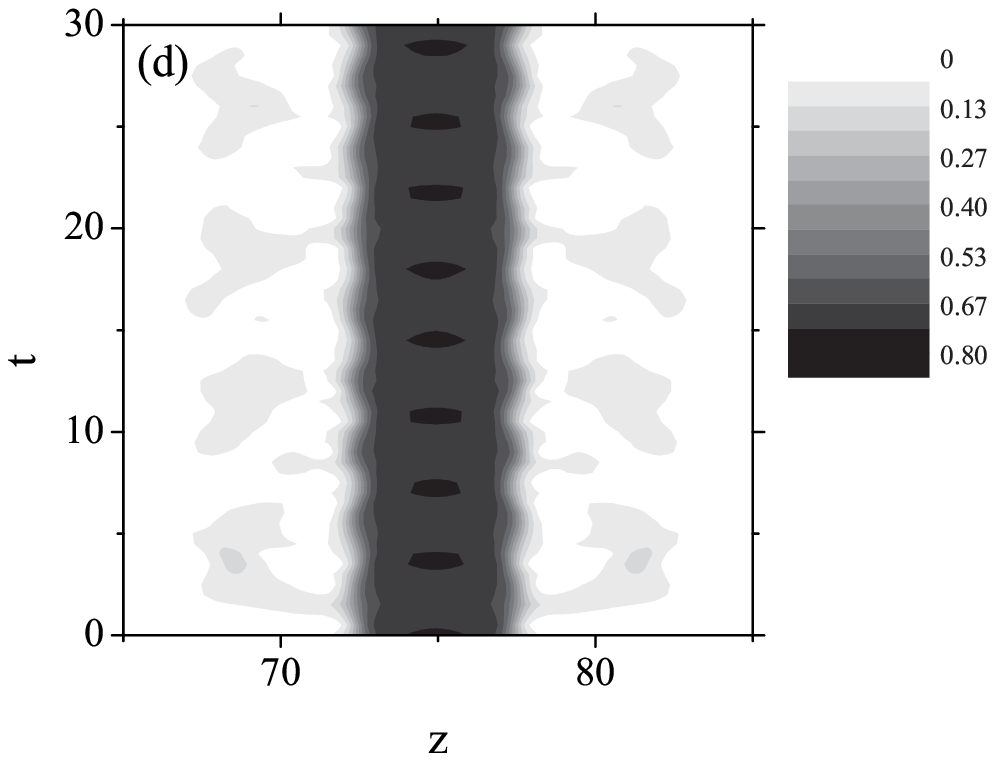}
\caption{(Color online) Profiles of the trapped modes in the
binary BEC with $\protect\kappa =0.1$, produced by the
imaginary-time simulations which were initialized by separated (a)
or overlapping (b) Gaussians. In the absence of the DD
interactions, the corresponding state is immiscible, independently
of the initialization. In the presence the DD interaction, the
modes evolve into breathers, as plotted for both components in (c)
and (d) for $\Gamma =40$ [the respective initial profile is
presented
by the dashed line in (b)]. Parameters of the DD interactions are $%
G_{11}=\Gamma =40,\,G_{21}=G_{12}=1.1\Gamma ,$ and
$G_{22}=1.21\Gamma $. } \label{fig8}
\end{figure}

Here we present results for slightly different magnitudes of the
dipole moments, \textit{viz}., $G_{11}\equiv \Gamma
,G_{22}=1.21\Gamma ,G_{21}=G_{12}=1.1\Gamma $, and several
different values of $\Gamma $. In the presence of the repulsive DD
interactions at small $\kappa $, immiscible states are formed, in
which the shapes of the two components may be spatially
asymmetric. Depending on the initial profile used in the
imaginary-time simulations, which were taken as two Gaussians with
coinciding or separated centers, two distinct types of stationary
immiscible states could be found, respectively, in the binary
system: those which feature the mirror symmetry between $\phi
_{1}(z)$ and $\phi _{2}(z)$, and asymmetric ones, see Fig.
\ref{fig8}. Computing energies of the observed states, we have
found that the energy is smaller (by a few per cent) in the
asymmetric state. Nevertheless, it does not represent the ground
state of the system, because in direct simulations this stationary
state develops into a breather, as shown in Fig. \ref{fig8}. Note
that the component with the larger dipole moment features a
smaller amplitude. Opposite to the case of equal dipole moments,
where the growth of the repulsive DD interaction could lead to the
IMT at $\kappa \neq 0$, here the growth of the DD repulsion
increases the separation between the components. A noteworthy
difference of the immiscible states in the present case from their
counterparts in the model with equal dipole moments (cf. Figs. \ref{fig1}, %
\ref{fig2}, \ref{fig3} for the case of the co-polarized components, and Fig. %
\ref{fig6} for the counter-polarization) is that the unequal
moments give rise to \textit{three-lobe} shapes. They feature two
regions occupied by the component with the larger dipole moment,
which are separated by a single region in which the component with
the smaller moment is concentrated.

The character of the mixed states generated due to the linear coupling is
not changed in the presence of the repulsive DD (of course, the shapes of
the two components become different). Real-time simulations indicate that
the immiscible and miscible localized states in the binary system with
different dipole moments of the components evolve into breathing structures,
which tend to keep the original immiscible or miscible structure of the
pattern. This is the case for both symmetric and asymmetric profiles of the
two components, which is understandable due to very close energies of the
immiscible and miscible modes (the energy difference between them is $%
\lesssim 1\%$). An example of the breather developing from immiscible modes
is displayed in Figs. \ref{fig8}(c,d).

\begin{figure}[tbp]
\center\includegraphics[width=5.7cm]{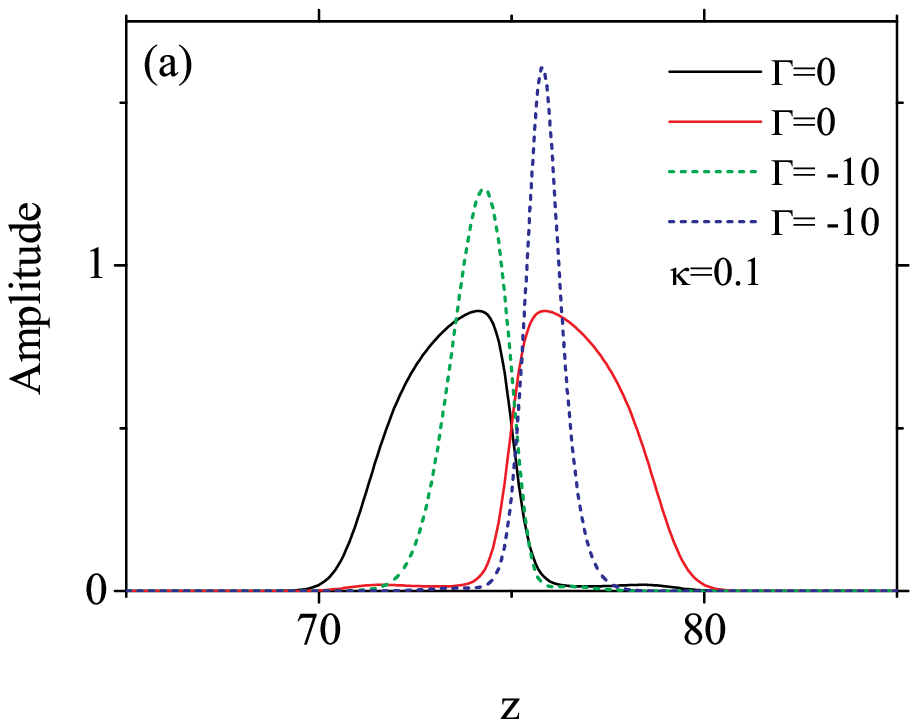}%
\includegraphics[width=5.7cm]{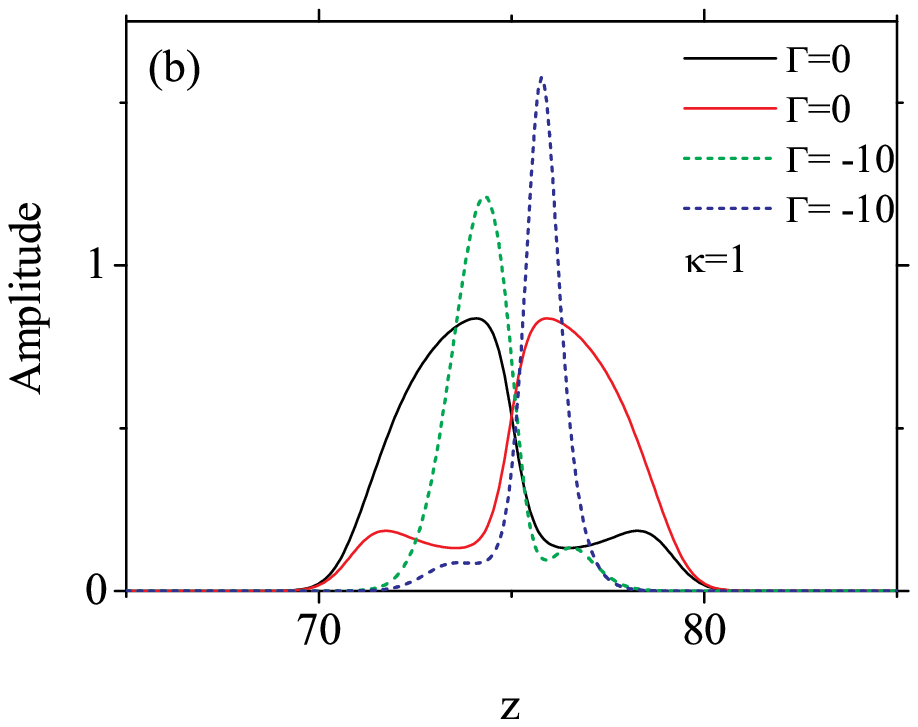} \includegraphics[width=5.7cm]{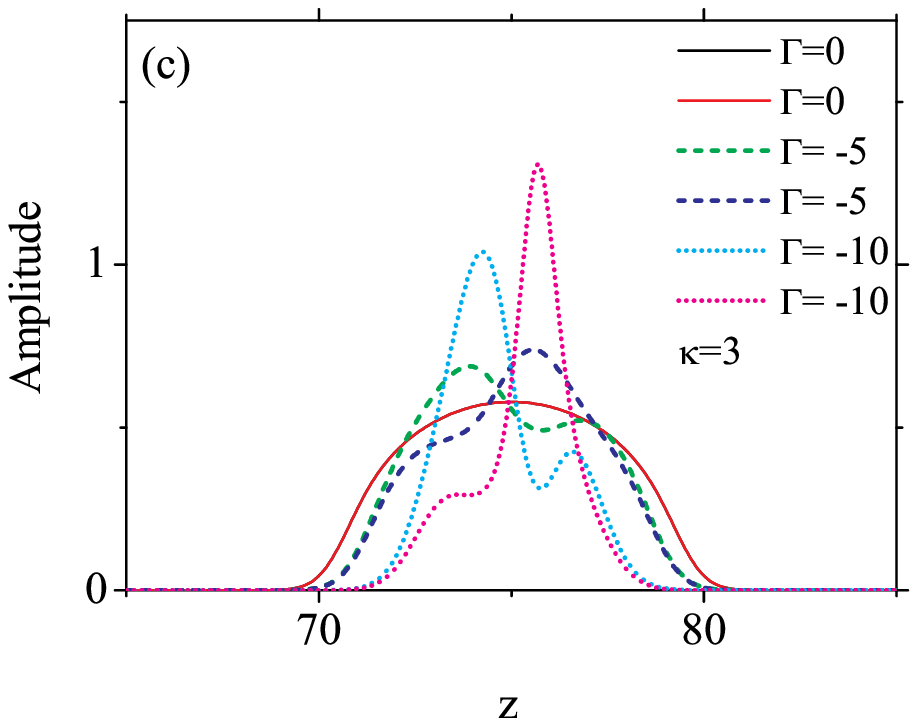}
\caption{(Color online) Profiles of asymmetric stationary states
in the
binary BEC for different values of the linear-coupling constant: (a) $%
\protect\kappa =0.1$, (b) $\protect\kappa =2$, and (c)
$\protect\kappa =3$. The strengths of the attractive DD
interaction are $G_{11}=\Gamma
,G_{22}=1.21\Gamma,$ and $G_{21}=G_{12}=1.1\Gamma $. The respective values of $%
\Gamma $ are indicated in the plots.}
\label{fig9}
\end{figure}

\begin{figure}[tbp]
\center\includegraphics [width=8cm]{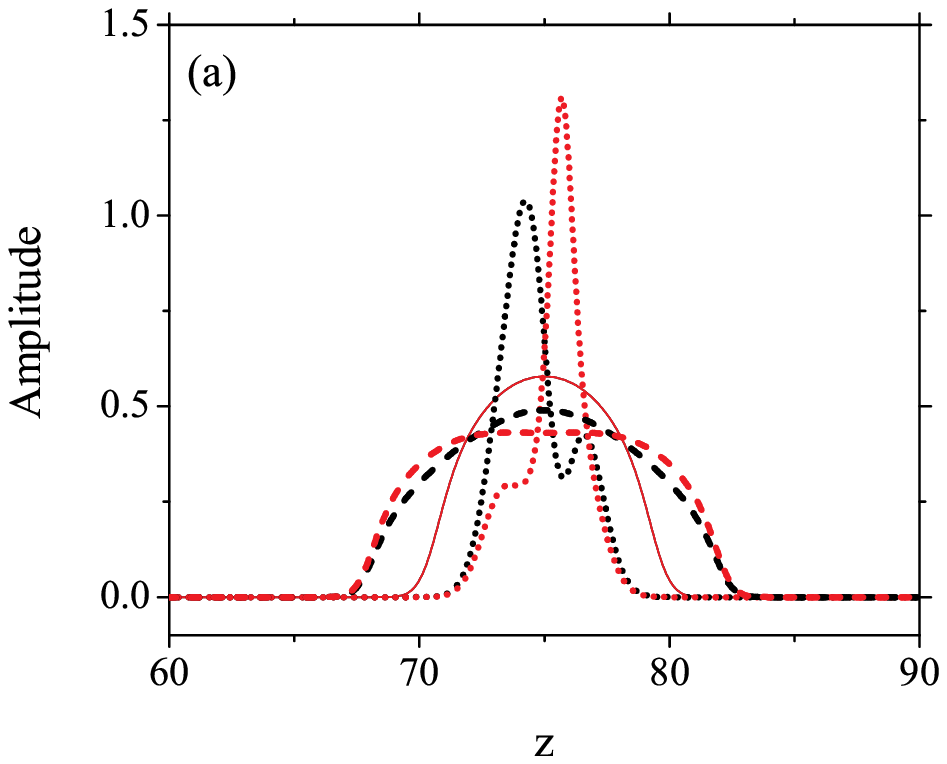}%
\includegraphics
[width=8cm]{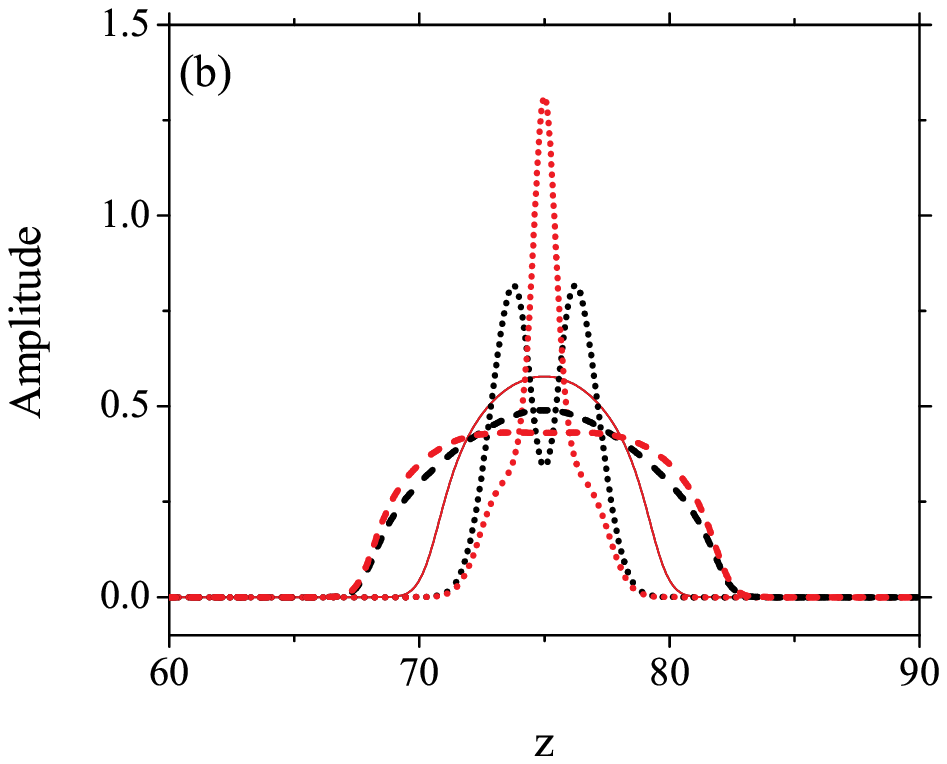} \includegraphics [width=8cm]{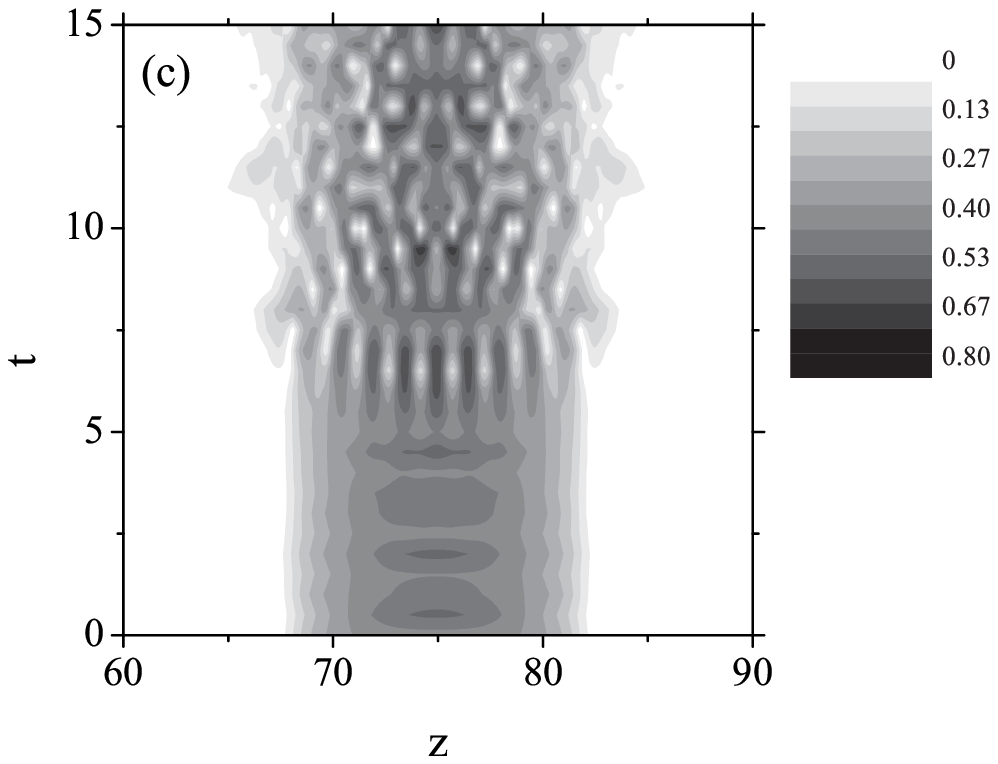}%
\includegraphics
[width=8cm]{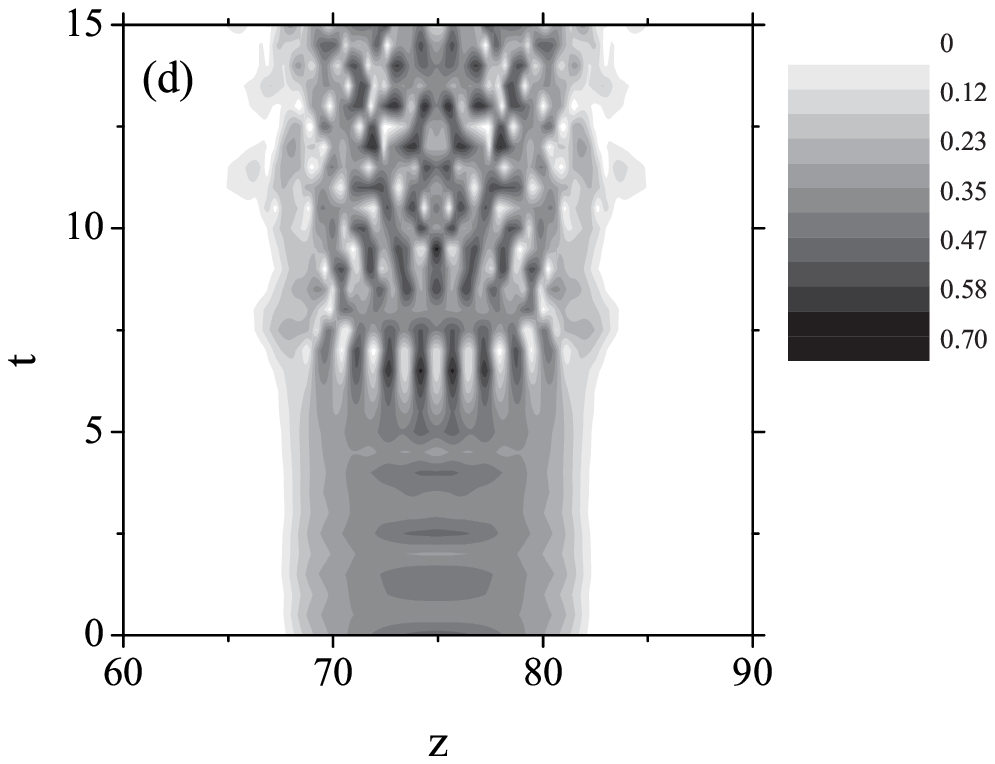} \includegraphics
[width=8cm]{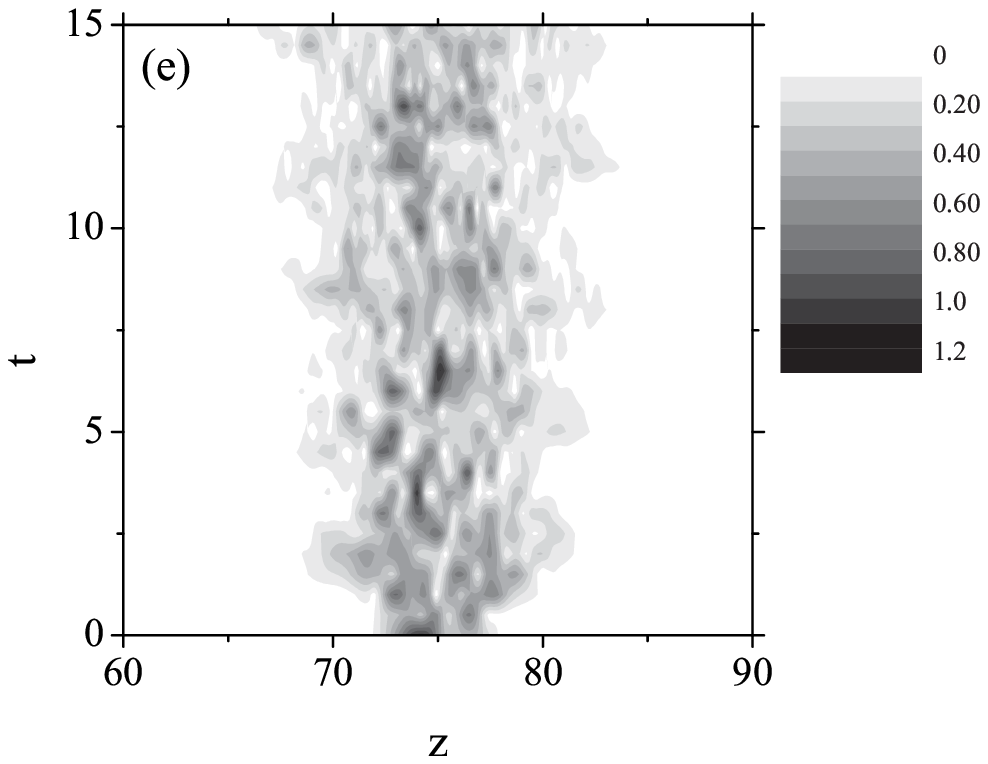}\includegraphics
[width=8cm]{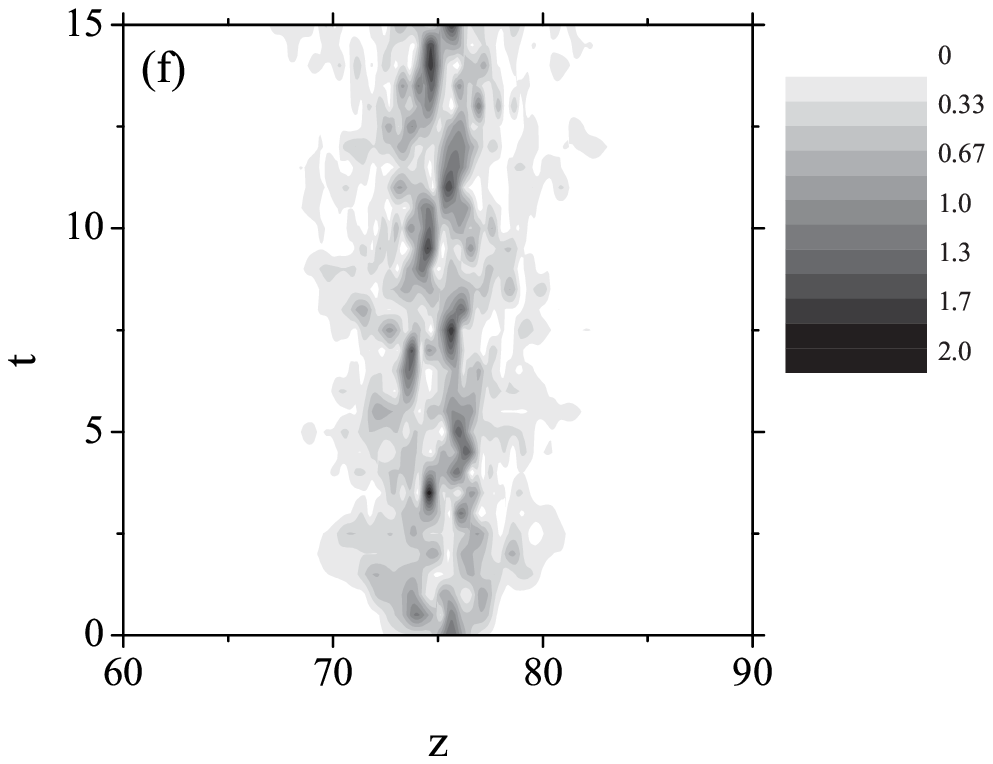}
\caption{(Color online) Profiles of the localized modes in binary BEC with $%
\protect\kappa =3$, obtained by means of the imaginary-time simulations with
initial conditions in the form of separated (a) and overlapping (b)
Gaussians: $\Gamma =0$ (solid lines), $\Gamma =40$ (dashed lines), and $%
\Gamma =-10$ (dotted lines). In the absence of the DD interaction, the
corresponding state is miscible, irrespective of the initialization. The
states are unstable for either sign of the DD interaction, as shown in (c),
(d) for $\Gamma =40$ [the respective initial profile is shown by the dashed
line in (b)] and in (e), (f) for $\Gamma =-10$ [the initial profile is
presented by the dotted line in (b)]. In both cases, the twin panels display
the evolution of the two components. Parameters of the DD interactions are $%
G_{11}=\Gamma ,\,G_{21}=G_{12}=1.1\Gamma,$ and $G_{22}=1.21\Gamma
$. } \label{fig10}
\end{figure}

The attractive DD interaction pushes all binary states in the system with
unequal dipole moments to enhanced immiscibility, see Fig. \ref{fig9} (which
is consistent with effects of the DD attraction outlined above in the
symmetric system). As well as in the case of the DD repulsion, the symmetry
of the stationary states obtained from the imaginary-time simulations
depends on the initialization, cf. Figs. \ref{fig8} and \ref{fig10}(a),(b).
The separated asymmetric states are getting more and more asymmetric and
both symmetric and asymmetric ones are becoming more and more narrow with
the growth of $|\Gamma |$. The maximum energy difference between the
symmetric and asymmetric states for fixed $\Gamma $ is of the order of a few
per cents. The stationary mixed state which exists without the DD
interaction for certain values of $\kappa $ is transformed by the DD
attraction into immiscible states whose symmetry is determined by the
initial conditions, see Fig. \ref{fig10}(b). Note that, in the case of the
DD attraction, the stationary localized states can only be found if the DD
interaction strength is not too large. Real-time simulations demonstrate
that all localized states which exist in the presence of the attractive DD
interaction are \emph{unstable}, see Figs. \ref{fig10} (c)-(f).

The results presented above, and additional ones not displayed
here, that were obtained for different combinations of the DD
interaction parameters, show a great diversity of possible
localized states in binary BEC with components formed of atom with
different moments. Generally, their symmetry properties, shape,
and width depend on the initial conditions.

\section{Conclusion}

In this work, we aimed to study effects of the DD interactions on
IMT  in binary BEC trapped in the HO potential. The interest to
this problem originates from the possibility to study the
interplay of the nonlocal DD attraction or repulsion with the
global spatial structure induced by the IMT. This binary system is
a complex one, as it depends on many parameters, such as the
orientation of the dipoles and the strength of the interaction
between them, numbers of atoms in each component, and the rate of
the linear interconversion between the components induced by the
external spin-flopping electromagnetic wave, in case the two
components represent two different spin states of the same atom.

We have reported the results for three dipolar binary BEC systems:
with co-polarized or counter-polarized dipolar components and
equal magnitudes of the dipole moments in the two components, and
for the system with unequal moments. Both repulsive and attractive
DD interactions were considered. In all the cases, the IMT
crucially depends on the presence of the linear coupling between
the two species, while the location of the miscibility threshold
and the stability of the resultant immiscible and miscible states
may be significantly affected by the DD interactions. In general,
the repulsive and attractive DD interactions shift the equilibrium
in favor of the miscibility and immiscibility, respectively (i.e.,
the long-range repulsion tends to suppress the spatial structure
induced by the immiscibility, while the nonlocal attraction helps
to enhance it, although with the trend to make it unstable against
the spontaneous transformation into a breather). In those cases
when regular breathers develop from unstable immiscible or
miscible stationary modes, they tend to keep the original
immiscible/miscible structure of the pattern. Only in the system
formed by the components with equal counter-polarized dipole
moments, the IMT may be induced by the DD interactions in the
absence of the linear interconversion between the constituents in
the binary BEC. On the other hand, in the case of the attractive
DD interactions, stable localized states do not exist for too
strong attraction. In particular, slightly perturbed miscible
narrow states, formed in the binary BEC with a very strong DD
attraction, are highly unstable and quickly disintegrate.

The patterns predicted in this analysis may be realized by means of the
available experimental techniques. In particular, the chromium condensate
may be created in a quasi-1D trap with the aspect ratio $\sim 10$, the axial
length about $10$ $\mathrm{\mu }$m, and the total \ number of atoms $\sim
25,000$ \cite{experim1,review}. This atom number and length should be
sufficient to allow the observation of the spatial structures expected in
the immiscible patterns considered above. The actual strength of the DD
interactions between the chromium atoms is equivalent to the effective
scattering length $\simeq 0.75$ nm, which can be made comparable to the
strength of the contact interactions if they are properly attenuated by
means of the Feshbach-resonance technique \cite{experim1}.

It may be interesting to extend the analysis reported in this work to a 2D
setting. It should also be relevant to extend the model by taking into
regard additional physical factors which are known to occur in dipole
condensates, such as atomic loss due to inelastic collisions \cite{review}.

\acknowledgments G.G., A.M., M. S., and Lj.H. acknowledge support from the
Ministry of Science, Serbia (Project 141034). The work of B.A.M. was
supported, in a part, by grant No. 149/2006 from the German-Israel
Foundation. This author appreciates hospitality of the Vin\v{c}a Institute
of Nuclear Sciences (Belgrade, Serbia).

\end{document}